\documentclass[preprint,12pt]{elsarticle}
\usepackage{aasmacros}

\usepackage{amssymb}
\usepackage{float}
\usepackage{threeparttable}
\usepackage{amsmath}
\usepackage{color}
\usepackage[hang,flushmargin]{footmisc}
\journal{Astroparticle Physics}

\begin{document}

\begin{frontmatter}



\title{Estimate of the Background and Sensitivity of the Follow-up X-ray Telescope onboard Einstein Probe}


\author[1]{Juan Zhang\corref{cor1}}%
\ead{zhangjuan@ihep.ac.cn}
\cortext[cor1]{Corresponding author}

\affiliation[1]{organization={Key Laboratory of Particle Astrophysics, Institute of High Energy Physics},%Department and Organization
            addressline={Chinese Academy of Sciences}, 
            city={Beijing},
            postcode={100049}, 
            %state={},
            country={China}}

\author[1]{Liqiang Qi}
\author[1]{Yanji Yang}
\author[1]{Juan Wang}
\author[2]{Yuan Liu}
\author[1]{Weiwei Cui}
%\author[1]{\\D. Author}%
\author[2]{Donghua Zhao}
\author[1]{Shumei Jia}
\author[1,3]{Tianming Li}
\author[1]{Tianxiang Chen}
\author[1]{Gang Li}
\author[1]{Xiaofan Zhao}
\author[1]{Yong Chen}
\author[4]{Huaqiu Liu}
\author[2]{Congying Bao}
\author[1]{Ju Guan}
%\author[1]{Chengkui Li}
%\author[1]{Haisheng Zhao}
%\author[4]{Fei Xie}
%\author[5]{Helen Poon}
\author[1]{Liming Song}
\author[2]{Weimin Yuan}

\affiliation[2]{organization={National Astronomical Observatories},%Department and Organization
            addressline={Chinese Academy of Sciences}, 
            city={Beijing},
            postcode={100012}, 
            %state={State Two},
            country={China}}
            
\affiliation[3]{organization={University of Chinese Academy of Sciences},%Department and Organization
            addressline={Chinese Academy of Sciences}, 
            city={Beijing},
            postcode={100049}, 
            %state={},
            country={China}}
            
\affiliation[4]{organization={Innovation Academy for Microsatellites},%Department and Organization
            addressline={Chinese Academy of Sciences}, 
            city={Shanghai},
            postcode={201210}, 
            %state={State Two},
            country={China}}

\begin{abstract}
As a space X-ray imaging mission dedicated to time-domain astrophysics, 
the Einstein Probe (EP) carries two kinds of scientific payloads, the wide-field X-ray 
telescope (WXT) and the follow-up X-ray telescope (FXT). 
FXT utilizes Wolter-I type mirrors and the pn-CCD detectors.
In this work, we investigate the in-orbit background of FXT 
based on Geant4 simulation. 
The impact of various space 
components present in the EP orbital environment are considered, such as 
the cosmic photon background, cosmic ray primary and secondary particles
(e.g. protons, electrons and positrons), albedo gamma rays, and 
the low-energy protons near the geomagnetic equator. 
The obtained instrumental background at 0.5-10\,keV, which is mainly induced by cosmic ray protons 
and cosmic photon background, corresponds to a level of 
$\sim$3.1$\times$10$^{-2}$ counts\,s$^{-1}$\,keV$^{-1}$
in the imaging area of the focal plane detector (FPD), i.e. 
3.7$\times$10$^{-3}$ counts\,s$^{-1}$\,keV$^{-1}$\,cm$^{-2}$ after 
normalization.
Compared with the instrumental background, 
the field of view (FOV) background, 
which is induced by cosmic photons reflected by the optical mirror, dominates below 2\,keV.
Based on the simulated background level within the focal spot 
(a 30$^{\prime\prime}$-radius circle), 
the sensitivity of FXT is calculated, 
which could theoretically achieve several $\mu$crab (in the order of 10$^{-14}$\,erg\,cm$^{-2}$\,s$^{-1}$) in 0.5-2\,keV 
and several tens of $\mu$crab (in the order of 10$^{-13}$\,erg\,cm$^{-2}$\,s$^{-1}$) in 2-10\,keV 
for a pointed observation with an exposure of 25 minutes. 
This sensitivity becomes worse by a factor of $\sim2$ if 
additional 10\% systematic uncertainty of the background 
subtraction is included.
\end{abstract}




\end{frontmatter}


\section{Introduction}\label{sec::intro}
 
The Einstein Probe (EP) is a space X-ray astronomical mission led by the Chinese Academy of Sciences (CAS) 
in collaboration with the European Space Agency (ESA) and the Max-Planck-Institute for extraterrestrial 
Physics (MPE), Germany. It is scheduled to be launched into a low-Earth orbit (LEO) with an altitude 
of $\sim$600\,km and an inclination of 29$^\circ$ by the end of 2022. 
It is dedicated to time-domain astrophysics with primary goals to discover 
cosmic high-energy transients and monitor variable objects \cite{2018SPIE10699E..25Y,2016SSRv..202..235Y,2018SSPMA..48c9502Y}. 
%To discover transients and monitor variable objects through the X-ray sky 
%with wide field and high sensitivity, 
EP carries two kinds of scientific payloads, the Wide-field X-ray Telescope (WXT, 0.5-4\,keV) and 
the Follow-up X-ray Telescope (FXT, 0.3-10\,keV). WXT will be used to
capture transients and to monitor variable objects, whereas 
FXT will be used to conduct deep follow-up observations 
of interesting targets discovered by WXT as well as by other facilities. 
WXT is equipped with the micro-pore lobster-eye optics,  
which achieves a large instantaneous field-of-view (FOV), 
3600 square degrees ($\sim$1.1\,sr), 
and a high sensitivity that is at least one order of magnitude better than 
those of the coded mask telescopes and pinhole/slit cameras currently in 
orbits, such as Swift/BAT and MAXI 
%with similar geometry design or physical dimensions 
\cite{2017ExA....43..267Z}. 
FXT uses Wolter-I optics with a focal length of 1.6\,m
%, to gathering X-rays from 0.5\,keV to 10\,keV, 
and the pn-CCD based focal plane detector (FPD) with pixel size of 
75\,$\mu$m$\times$75\,$\mu$m \cite{2018SSPMA..48c9502Y}. 
%The background and sensitivity of FXT have not been investigated before.

To estimate the in-orbit background of detectors of space instruments before launch 
is an essential part of any X-ray mission development. 
On the one hand, these studies can be used not only to optimize the instrument design and  
to evaluate the onboard storage requirements, but also 
to investigate whether the instrument could fulfill the scientific goals. 
On the other hand, the estimated background helps scientists to 
understand the performance of the telescope and 
conduct the pre-study of target sources of interest, 
and helps the mission operation team to draw up observation strategies. 
Each space instrument needs to conduct its own background 
estimation due to the different detector type and operation environment 
for different orbits, 
e.g. Tenzer et al. \cite{2010SPIE.7742E..0YT}, Perinati et al. \cite{2012ExA....33...39P} and 
Weidenspointner et al. \cite{Weidenspointner:2008oka} for eROSITA, 
Fioretti et al. \cite{2016SPIE.9905E..6WF} for ATHENA, 
Campana et al. \cite{2013ExA....36..451C} for LOFT/LAD, 
Xie and Pearce \cite{2018Galax...6...50X} for Sphinx, 
Xie et al. \cite{2015Ap&SS.360...47X} and Zhang et al. \cite{2020Ap&SS.365..158Z} for Insight-HXMT, and 
Zhao et al. \cite{2017ExA....43..267Z} for a micro-pore lobster-eye telescope which is the prototype of EP/WXT. 

In this work, we %intend to 
investigate the background and sensitivity 
of FXT. A method to estimate the in-orbit background of space instrument is to use 
Monte Carlo simulations facilitated by the
Geant4 toolkit \cite{Agostinelli:2002hh,Allison:2006ve,Allison:2016lfl}, 
which was developed for simulation of particle physics experiments. 
Geant4 could depict the interaction of particles with matter as realistically as possible. 
Different with what done 
for Insight-HXMT \cite{2015Ap&SS.360...47X,2020Ap&SS.365..158Z}, the 
simulation of FXT needs the process that describes the reflection 
of photons incident on the Wolter-I mirror with grazing angles. 
Currently Geant4 lacks 
these processes in its release versions. Fortunately, extended 
packages have been developed that could be incorporated into Geant4,
e.g. the generic ray-tracing extension XRTG4 developed by 
Buis and Vacanti \cite{2009NIMPA.599..260B} and 
the physics model for the interaction of X-rays and matter 
at grazing incidences developed by Qi et al. \cite{2020NIMPA.96363702Q}. 
For the proton scattering at grazing incidence angles off X-ray mirrors, 
the analytical model and semi-empirical model (e.g. \cite{1980JETP...52..225R,2020ExA....49..115A}) and Geant4 packages 
(e.g. \cite{2020NIMPA.96363702Q,2017ExA....44..401G} and references therein) have been developed. 
In this work, the background simulation of FXT is conducted 
with Geant4 and the external packages developed by Qi et al.  \cite{2020NIMPA.96363702Q}. 
The mass modeling of FXT is established, including EP platform, 
WXT and FXT. The space radiation components in EP orbit environment 
are widely investigated. 
Based on the simulated background, the sensitivity of FXT is investigated.

\section{EP/FXT}
\label{sec::fxt}

\begin{figure}[H]
\centering
\includegraphics[width=0.5\columnwidth]{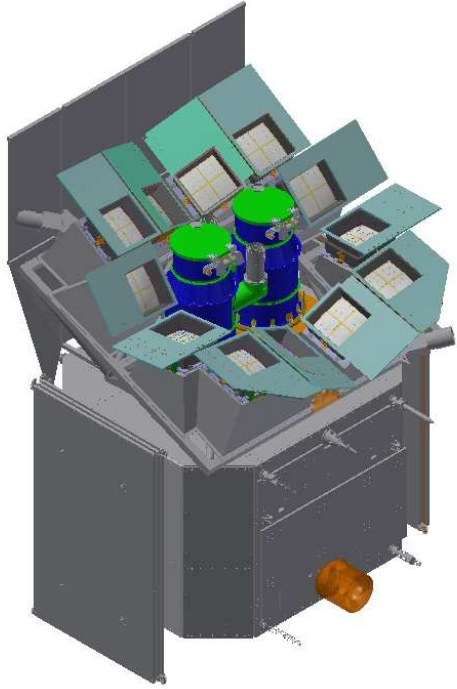} % epweb.eps
\caption{The latest configuration of EP. 
The two blue tubes in the center are the two FXT modules. 
Surrounded are the 12 modules of WXT.
}
\label{fig::epsat}
\end{figure}

Figure\,\ref{fig::epsat} shows the latest configuration of EP.
It consists of two scientific payloads, WXT and FXT.
WXT is equipped with lobster-eye optics and has 12 identical 
modules forming a total FOV of 3600 square 
degrees ($\sim$1.1\,sr), which makes it have excellent grasp 
capability \cite{2018SSPMA..48c9502Y} and ideal for the whole sky 
monitoring of X-rays. FXT, as the follow-up observation telescope,
utilizes Wolter-I mirrors and pn-CCD detectors forming a circular FOV 
with a diameter of $\sim$1 degree ($3\times10^{3}$\,arcmin$^{2}$). 
Compared to WXT, FXT has much larger effective area and 
angular resolution \cite{2018SPIE10699E..25Y,2016SSRv..202..235Y,2018SSPMA..48c9502Y}. 
FXT comprises two co-aligned identical modules, %which are 
surrounded by the 12 modules of WXT. 
The development of FXT is a collaboration among the CAS, ESA and MPE. 
The Institute of High Energy Physics (IHEP), CAS is responsible for the overall design, 
development and test of the entire FXT instrument, while ESA and MPE contribute to the 
FXT development via provision of some of the key components of FXT including the mirror assemblies, 
use of the mirror design and mandrels, electron diverter, and 
CCD detector modules\footnote{ESA provides one set of the Mirror 
Assembly and the Electron Diverter, and MPE provides the eROSITA 
design information and use of the mandrels, one eROSITA Mirror Flight 
Spare and Mirror Demonstrator Model, and a number of detector modules plus CAMEX test module.}. 

\begin{figure}[H]
\centering
\includegraphics[width=0.5\columnwidth]{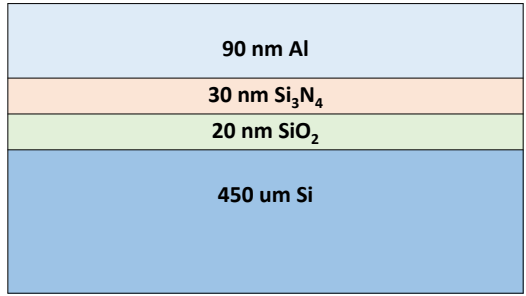}
\caption{On-chip filters above the pn-CCD. 
}
\label{fig::pnccd}
\end{figure}

The optics of FXT is similar to that of the eROSITA telescopes, 
which consists of 54 Wolter-I type mirror shells, 
with an outer diameter of $\sim$36\,cm on the top and 
7.6\,cm at the bottom and a  focal length of 1.6\,m. 
For each shell, a layer of 100\,nm thick gold is coated on the 
nickel substrate. The FPD of FXT is the pn-CCD.
The sensitive layer of the pn-CCD is 450\,$\mu$m silicon, 
which is covered by 20\,nm SiO$_{2}$, 30\,nm Si$_3$N$_4$ and 
an on-chip filter of 90\,nm aluminum, as illustrated in Figure\,\ref{fig::pnccd}.
The FPD has an imaging area and a frame-store area. 
There are $384\times384$ pixels in the imaging area, 
each pixel with a size of 75\,$\mu$m$\times$75\,$\mu$m. 
The frame-store area also has $384\times384$ pixels correspondingly, 
but with smaller pixel size, 51\,$\mu$m$\times$75\,$\mu$m. 
The FPDs of the two FXT modules are placed by rotating 
an angle of 90 degree relative to each other in the focal plane of FXT 
so that they could be complementary to each other 
during observation. 
The pn-CCD is placed inside an aluminum alloy detector box, 
which is nested into a $\sim$3\,cm thick copper box, 
in order to shield the cosmic ray particles and photon background 
in the space environment outside the FOV of FXT.

\begin{figure}[H]
\centering
\includegraphics[width=0.9\columnwidth]{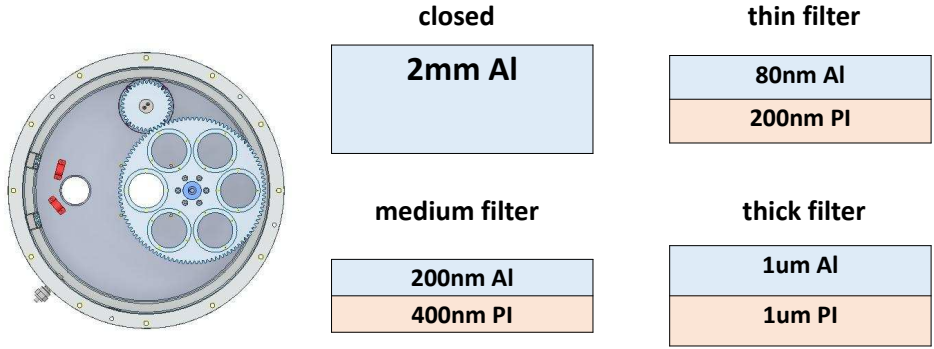}
\caption{The filter wheel of FXT.}
\label{fig::filterwheel}
\end{figure}

In addition to the on-chip filter, there is also a filter wheel 
between the Wolter-I optics and the FPD inside each FXT module. 
There are six positions on the filter wheel, as illustrated in Figure\,\ref{fig::filterwheel}. 
The closed position is filled with a 2\,mm aluminum and is 
used to obtain the instrumental background during the in-orbit observation. 
The thin, medium and thick filters, which are different in the thickness
of the aluminum (Al) and the polyimide (PI), are 
used to observe sources with different fluxes to avoid the 
pile-up effect.
The other two positions, not shown in this figure, 
are the open setup and the in-orbit calibration source setup.

\section{Space environment}
\label{sec::spcenv}
EP will be launched into a circular LEO 
with an altitude of 600\,km and an inclination of 29$^\circ$ 
by the end of 2022. 
In such an orbit, there are various space radiation 
components that may cause background events in FXT, 
such as the cosmic photon background, 
the cosmic rays (primary and secondary protons, electrons,  
and positrons, etc.), the low energy charged particles 
which are enhanced near the geomagnetic equator, and the 
albedo gamma rays.
In this section, we describe these space radiation 
components 
and the models we adopt for the background simulation of FXT.

\subsection{Cosmic photon background}

The cosmic photon background includes emissions from the 
unresolved sources outside our Galaxy, i.e. the cosmic 
X-ray background (CXB), and the diffuse Galactic foreground.
Generally CXB is modeled by a broken power-law 
\cite{1992NIMPA.313..513G,1999ApJ...520..124G} 
from keV to several hundred of MeV. These models have 
been used in the simulation of LOFT/LAD \cite{2013ExA....36..451C} 
and Insight-HXMT \cite{2015Ap&SS.360...47X,2020Ap&SS.365..158Z}. 
For the diffuse Galactic foreground, the contributed photons 
are softer. 
McCammon et al. have given a high spectral resolution observation 
of the soft x-ray diffuse background, of which the spectrum could be fitted 
by a two-temperature thermal plus power-law model \cite{2002ApJ...576..188M}. 
This model is denoted by apec+wabs(apec+const*powerlaw) in XSPEC
and has been used for the background simulation of Athena/WFI\cite{wfibkg}. 
The parameters are listed in Table\,\ref{tab::softpara}.
The powerlaw term, wabs*const*powerlaw, 
represents the extragalactic diffuse emission from unresolved sources. 
The two-temperature thermal term, apec+wabs*apec, models the diffuse Galactic foreground. 

\begin{figure}[H]
\centering
\includegraphics[width=0.9\columnwidth]{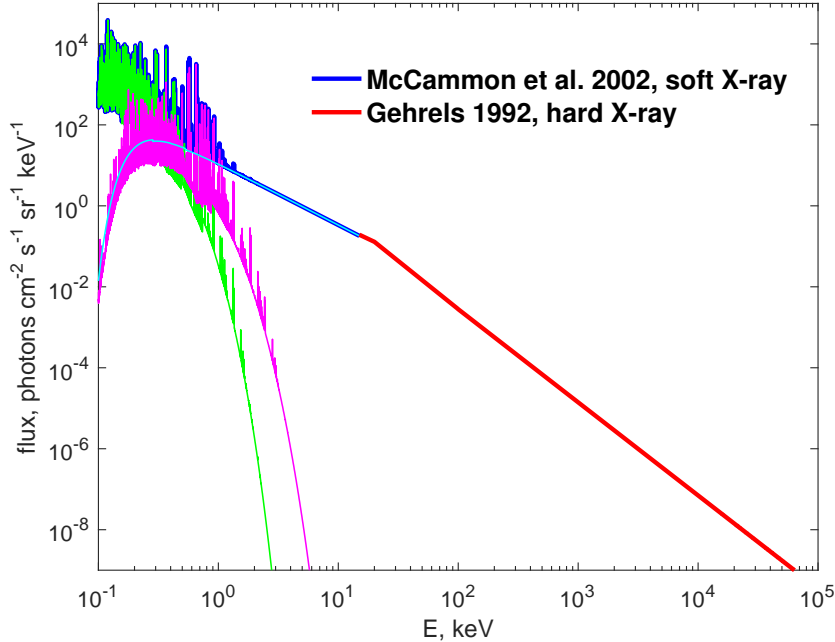}
\caption{The cosmic photon background spectra used in the FXT background 
simulation. The CXB spectrum \cite{1992NIMPA.313..513G} above 15\,keV 
is plotted in red, and labeled with ``Gehrels 1992, hard X-ray". The high 
resolution soft X-ray spectrum \cite{2002ApJ...576..188M} is denoted by 
``McCammon et al. 2002, soft X-ray". Also plotted are the three components of the 
soft spectrum. The two temperature equilibrium thermal emission with 
and without absorption are in magenta and green respectively. 
The absorbed power-law spectrum with an constant factor of 1 is 
plotted in cyan.}
\label{fig::input_cxb}
\end{figure}

Given the energy band of interest of FXT, 0.3-10\,keV, 
in this work the soft X-ray diffuse background spectrum 
\cite{2002ApJ...576..188M,wfibkg} is used for photons with 
energy less than 15\,keV and the CXB spectrum \cite{1992NIMPA.313..513G} 
is adopted above 15\,keV, as shown in Figure\,\ref{fig::input_cxb}. 
Note that the constant factor of the power-law component of the
soft X-ray spectrum model is taken as 1 for the background simulation of FXT  
instead of a value of 0.2 used for Athena/WFI. 
This is because the goal sensitivity of Athena/WFI is much better, 
$\sim 10^{-17}$ erg cm$^{-2}$ s$^{-1}$ for a 
100\,ks observation\footnote{https://tuprints.ulb.tu-darmstadt.de/2980/}. 
With such a good sensitivity, it is assumed that 80\% of the pointed 
sources could be resolved from CXB. Therefore a factor of 0.2 
is used for Athena/WFI \cite{wfibkg}.
While for EP/FXT, the typical pointing observational exposure is 1500 seconds 
according to the planned observational strategy. It doesn't have good enough 
sensitivity to resolve point sources from CXB. Thus a factor of 1 is used. 
The CXB spectrum adopted 
from  \cite{1992NIMPA.313..513G} is presented in Equation\,\ref{eq::cxbspc}.
\begin{equation}
\label{eq::cxbspc}
%\begin{aligned}
F=\left\{\begin{array}{ll}
0.54\times E^{-1.4}, & E<0.02\,{\rm MeV}\\
0.0117\times E^{-2.38}, & 0.02\,{\rm MeV}<E<0.1\,{\rm MeV}\\
0.014\times E^{-2.3}, & E>0.1\,{\rm MeV}
%\end{aligned}
\end{array}
\right.
\end{equation}
where {\it E} is in MeV and {\it F} in photons cm$^{-2}$ s$^{-1}$ MeV$^{-1}$ sr$^{-1}$.
At first during the simulation, we ignore the spacial anisotropy of the cosmic photon background
and take it to be uniformly distributed in the whole sky. 
This may be not suitable for soft X-rays diffuse background due to its change with different Galactic pointing directions. 
In the following sections, the impact of this change and its impact on  
the background and sensitivity will be discussed.

\begin{table}[H]
\caption{Model parameters for the soft X-ray diffuse background, which 
is expressed by wabs(apec+const*powerlaw) using XSPEC notation.
Normalizations are scaled to 1 arcmin$^2$. This table is adopted 
from \cite{wfibkg}. The difference is that the constant factor of 
the power-law component is taken as 1 for the background simulation of FXT}
\label{tab::softpara}
\centering
\begin{threeparttable}
\begin{tabular}{cccccccccc}
\hline
& Model & Parameter & Value &   &  &   &  &  & \\
\hline
&apec & kT\tnote{1)} & 0.099  \\
&apec & abundance &  1  \\
&apec & redshift  &  0 \\
&apec & norm\tnote{2)}  & 1.7E-6 \\
\hline
&wabs & NH\tnote{3)} & 0.018 \\
&apec & kT\tnote{1)}  & 0.225 \\
&apec & abundance &  1  \\
&apec & redshift &  0 \\
&apec & norm  & 7.3E-7 \\
&const &       & 1     \\
&powerlaw & photon index &  1.52  \\
&powerlaw & norm & 9.4E-7\tnote{4)} \\
\hline
\end{tabular}
\begin{tablenotes}
\item[1)] kT is in  keV
\item[2)] in  units of 
$10^{-14}/(4\pi D_{A}(1+z))^{2} \int n_{e}n_{H}dV$, where 
$D_{A}$ is the angular size distance to the source (cm), and n$_{e}$ and n$_{H}$ are the electron and H densities (cm$^{-3}$)
\item[3)] in units of $10^{22}$ cm$^{-2}$
\item[4)] in units of photons/keV/cm2/s @ 1 keV
\end{tablenotes}
\end{threeparttable}
\end{table}

\subsection{Cosmic rays}

\begin{figure}[H]
\centering
\includegraphics[width=0.7\columnwidth]{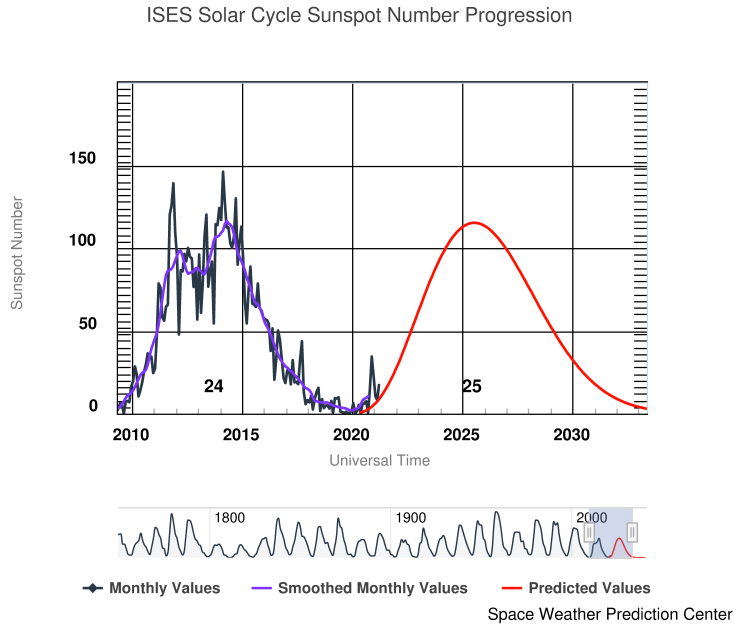}
\caption{The predicted solar activity. This figure is adopted from  https://www.swpc.noaa.gov/products/solar-cycle-progression}
\label{fig:input_crp}
\end{figure}

Cosmic rays are high energy charged particles generated from outside the Earth. 
Generally, it is believed that cosmic rays above $\sim$EeV  
originate from outside the Galaxy, while those below this energy 
come from our Galaxy. 
For LEO satellites, the encountered cosmic rays are modulated 
by the solar activity and the geomagnetic fields, as described in Equation\,\ref{eq::crspc}, 
\begin{align}
%\begin{equation}
\label{eq::crspc}
F(E) = F_{\rm LIS}(E+|Z|e\Phi)  &\times \frac{(E+mc^2)^2-(mc^2)^2}{(E+mc^2+|Z|e\Phi)^2-(mc^2)^2} \nonumber\\
&\times \frac{1}{1+(R/R_{\rm cut})^{-r}}
%\end{equation}
\end{align}
where E, m and Z are the kinetic energy, the rest mass and the charge of 
the cosmic ray particle, respectively. The first term, $F_{\rm LIS}$, is the cosmic ray flux 
of the local interstellar (LIS) environment. 
The second term depicts the solar modulation on $F_{\rm LIS}$.
It is described by the ``force field approximation" model \cite{1968ApJ...154.1011G}, 
where $\Phi$ is solar modulation potential, which varies from $\sim$300\,MV 
for solar activity minimum to $\sim$1300\,MV for solar activity maximum 
\cite{2018AnGeo..36..555F}. During the solar minimum, more low energy 
cosmic rays could enter the solar system. On the contrary, they will 
be shielded due to the higher potential during the solar maximum.
The design lifetime of EP is 3 years, with a goal of 5 years.
According to the predicted solar activity, as shown in Figure\,\ref{fig:input_crp}, 
the operation duration of EP is approaching to or around the solar maximum year.

The third term of Equation\,\ref{eq::crspc} 
describes the modulation of the geomagnetic field, where 
$r=12$ for protons, and $r=6$ for electrons and positrons \cite{2004ApJ...614.1113M}.
R is the particle rigidity, $R=pc/Ze$, where $p$ is the momentum of this particle. 
$R_{\rm cut}$ is the cut off rigidity of a given location at the orbit of EP. 
This parameter represents the minimum energy needed by a primary cosmic 
ray particle to reach a given location at the Earth. It is 
related to the altitude and the geomagnetic latitude of this location 
and could be calculated by assuming a dipole geomagnetic field \cite{2004ApJ...614.1113M}, 
\begin{equation}
\label{eq::cr_formula}
R_{\rm cut} = 14.9 \times \left( 1+ \frac{h}{R_{\rm Earth}} \right) ^{-2} ({\rm cos} \theta_{M})^4 {\rm GV}
\end{equation}
where $R_{\rm Earth}$ is the radius of the Earth, $h$ the space altitude, 
and $\theta_{M}$ the geomagnetic latitude. 
For the EP orbit, with an altitude of 600\,km and an inclination 
of 29\,degree, $\theta_{M}$ varies from 0 to 0.7 rad
\footnote{http://wdc.kugi.kyoto-u.ac.jp/igrf/gggm/index.html\\
https://omniweb.gsfc.nasa.gov/vitmo/cgm.html}. It can be seen that with the same altitude
in the EP orbit, the high geomagnetic latitude 0.7\,rad corresponds to 
a cut off rigidity 
of $\sim$4.3\,GV and the equatorial geomagnetic latitude corresponds to $\sim$12.6\,GV, which implies that there are more primary cosmic ray particles at several GeV 
around the high latitude of the geomagnetic field.

\subsubsection{Primary protons}
\begin{figure}[H]
\centering
\includegraphics[width=0.9\columnwidth]{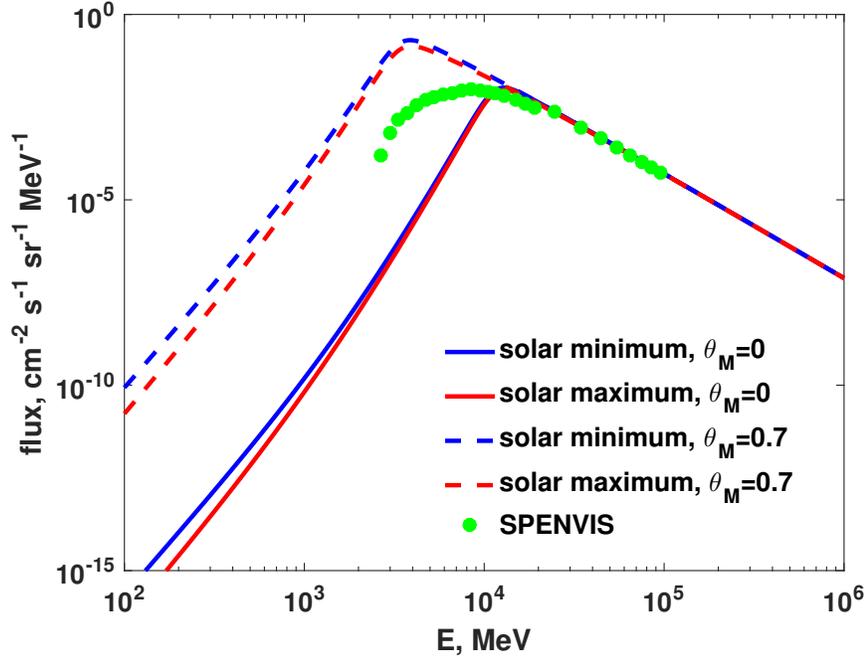}
\caption{
The spectra of primary cosmic ray protons in EP orbit. The blue and red lines 
denote the different solar activities, and the dashed and solid lines correspond 
to different geomagnetic latitudes. 
The spectrum shown as green dots are the average spectrum 
obtained from SPENVIS for the EP orbit.
}
\label{fig::primary_crp}
\end{figure}

Protons are the dominant component among cosmic ray particles.
% till here
The spectrum of primary cosmic ray protons has been precisely measured by AMS-01 
\cite{Alcaraz:2000ks,Alcaraz:2000vp} and AMS-02 \cite{Aguilar:2015ooa}. And the 
power-law spectral model based on measurements, 
$F_{\rm LIS} = 23.9 \times \left [ \frac{R(E)}{\rm GV}  \right ]^{-2.83}$ particles m$^{-2}$ 
s$^{-1}$ sr$^{-1}$ MeV$^{-1}$, was also widely used for the simulation 
of space instruments, e.g. \cite{2013ExA....36..451C,2004ApJ...614.1113M}. 
One can also obtain the orbital averaged spectrum of cosmic ray protons 
from ESA's SPace ENVironment Information System\footnote{https://www.spenvis.oma.be} (SPENVIS, 
one can get the geographic and geomagnetic information of a spacecraft, the radiation environment and dose, etc. from SPENVIS).
In this work, the spectrum from SPENVIS is used for the background simulation 
of FXT, as plotted in Figure\,\ref{fig::primary_crp}. It is worth noting that 
this spectrum is also consistent with the spectrum 
obtained by averaging the AMS-01 measurements from 0 to 0.7 rad \cite{Alcaraz:2000ks}.
We show in Figure\,\ref{fig::primary_crp} the 
spectra resulting from the power-law model and 
including modulation by solar activity 
and the effect of geomagnetic latitude. 
%(also show the ams averaged spectrum in the figure?)

\subsubsection{Secondary Protons}
\begin{figure}[H]
\centering
\includegraphics[width=0.9\columnwidth]{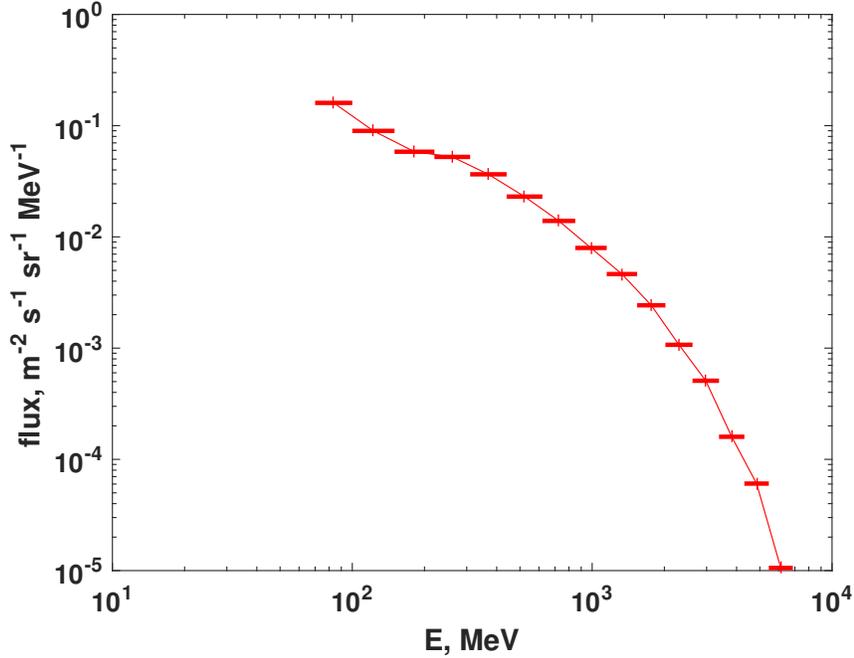}
\caption{The incident spectral model of secondary protons used in EP orbit.}
\label{fig::secondary_crp}
\end{figure}

Secondary cosmic ray particles are produced from the collisions of the primaries
with the Earth's atmosphere. They are also strongly dependent on the 
geomagnetic field. Their spectra in different geomagnetic cutoff can 
be modeled by the broken power-law or the cutoff power-law model \cite{2004ApJ...614.1113M}.
In this work, we use a spectrum obtained by averaging the AMS-01 measurements 
\cite{Alcaraz:2000ks} between $\theta_{M}=0\sim0.7$ rad
to simulate the background induced by secondary protons. 
This spectrum is shown in Figure\,\ref{fig::secondary_crp}.

\subsubsection{Low energy protons near the equator}
Besides the primary and secondary cosmic ray protons, 
it was found that there are enhancements of proton flux with energies 
from tens keV to several MeV near the geomagnetic equator \cite{1972ZGeo...38..701M}.
They are believed to be produced by the charge exchange between the inner radiation belt 
protons and the neutral atoms of the atmosphere, and the spectrum of these 
low energy protons can be modeled by a kappa function 
\cite{2008AdSpR..41.1269P,2009AdSpR..43..654P},
\begin{equation}
\label{eq::proton_kappa}
f(E)=A\left [ 1+ \frac{E}{kE_{0}}  \right ]^{-k-1},
\end{equation}
where A is in cm$^{-2}$ s$^{-1}$ sr$^{-1}$ keV$^{-1}$, E and E$_{0}$ in keV. 
A=50, k=2.3 and E$_{0}$=30 for the quiet geomagnetic conditions, and 
A=330, k=3.2 and E$_{0}$=22 for the disturbed conditions \cite{2009AdSpR..43..654P}.
Though they are easily blocked by the shielding materials around the 
detector, these low energy protons have to be taken into account 
due to the funneling effects of the focusing mirrors 
\cite{2000SPIE.4140...32K,2003ITNS...50.2018L,2020NIMPA.96363702Q,2021ExA...tmp...12Q}. 
In our simulation, the kappa function with A=328, k=3.2 and E$_{0}$=22 
\cite{2008AdSpR..41.1269P} is used.

\subsubsection{Primary electrons and positrons}
\begin{figure}[H]
\centering
\includegraphics[width=0.9\columnwidth]{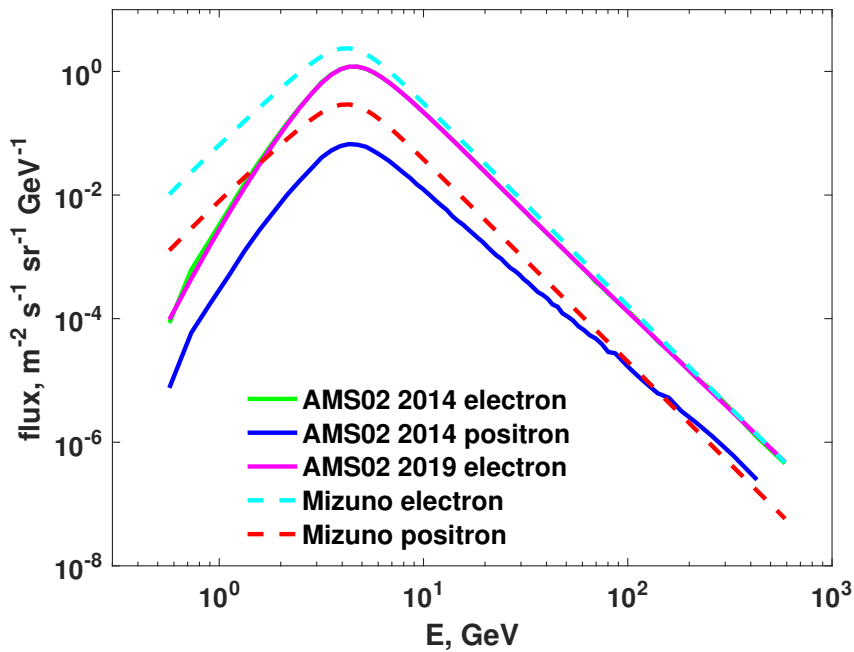}
\caption{
The spectra of primary cosmic ray electrons (dashed cyan) and 
positrons (dashed red) used for FXT simulation. Also plotted are 
the electron and positron spectrum of AMS-02 
\cite{2014PhRvL.113l1102A,2019PhRvL.122j1101A} after the geomagnetic modulation 
at $\theta_M=0.7$ in the EP orbit. 
}
\label{fig::cr_ep}
\end{figure}

The latest precise measurements of AMS-02 showed that the spectra of 
primary cosmic ray electrons and positrons have different magnitudes and energy dependences \cite{2014PhRvL.113l1102A,2019PhRvL.122j1101A}. 
Besides these newest measurements, a frequently used model of the LIS electrons 
is given by Mizuno et al. \cite{2004ApJ...614.1113M},
\begin{equation}
\label{eq::primary_e}
F_{\rm LIS} = 0.65 \times \left [ \frac{R(E)}{\rm GV}  \right ]^{-3.3} {\rm counts\, m^{-2} s^{-1} sr^{-1} MeV^{-1}}
\end{equation}
where R(E) is the rigidity of the electron. And the positron ratio, $e^{+}/(e^{+}+e^{-})$, is generally taken as 0.11 \cite{1994ApJ...436..769G}.
Figure\,\ref{fig::cr_ep} shows the modeled primary electrons and positron 
spectra for 
a solar minimum and a geomagnetic latitude of $\theta_M=0.7$. Also shown are 
the primary cosmic ray electrons and positrons of AMS-02  \cite{2014PhRvL.113l1102A,2019PhRvL.122j1101A} after modulation by the geomagnetic 
field at $\theta_M=0.7$. To obtain a conservative estimate of the background, 
the modeled spectra are used in this work.
%The spectra of all these models are shown in Figure\ref{fig::cr_ep}.

\subsubsection{Secondary electrons and positrons}

\begin{figure}[H]
\centering
\includegraphics[width=0.9\columnwidth]{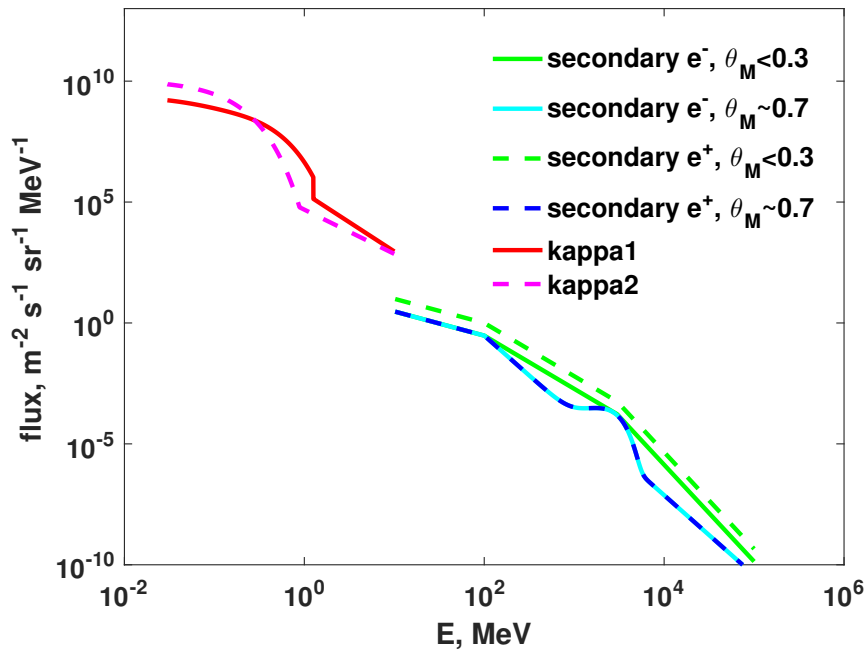}
\caption{Secondary electron and positron spectra at different 
geomagnetic latitudes. The two spectra below 10\,MeV are the 
enhanced electrons at $L<2$ \cite{2008AdSpR..42.1523G}.}
\label{fig::secondary_ep}
\end{figure}

The spectra of secondary electrons and positrons can be modeled by the PL model, 
the broken power-law model or a power-law model with a hump model, which is 
correlated with the geomagnetic latitudes \cite{2004ApJ...614.1113M}. 
Different from the primary electrons and positrons,
the positron/electron ratio varies with the geomagnetic latitude 
\cite{2004ApJ...614.1113M}. 
The secondary electron and positron spectra at regions of 
$0\leq\theta_M\leq0.3$ and $0.6\leq\theta_M\leq0.8$ are shown 
in Figure\,\ref{fig::secondary_ep}. 
For the simulation of FXT, we adopt the upper spectra corresponding 
to those at $0\leq\theta_M\leq0.3$. 

In Figure\,\ref{fig::secondary_ep} we show the spectra of low 
energy electrons from 10\,keV to 10\,MeV 
for the two flux-enhancement areas, i.e. the near-equatorial zone 
and the middle latitude zone \cite{2008AdSpR..42.1523G}. 
For Wolter-I type X-ray telescopes, generally there is a magnetic deflector 
at the back end of the mirror to prevent low-energy charged particles 
from focusing on the FPD. In fact, low-energy electrons could be completely
shielded with certain configurations of the magnetic deflector \cite{2021ExA...tmp...12Q}. 
FXT is also equipped with the magnetic deflector. 
Therefore these low-energy electrons are not included in the following background simulation.

\subsection{Albedo gamma rays}

The albedo gamma rays originate from the interaction of the cosmic rays  
with the atmosphere and the atmosphere reflection of CXB. 
Therefore the albedo gamma ray flux strongly depends on the  
direction of the Earth relative to the instrument FOV.
In this work, the spectrum of albedo gamma rays from Campana et al. 
\cite{2013ExA....36..451C} is adopted. According to the current observing strategy of 
EP\footnote{http://ep.nao.cas.cn/epmission/epoperation/201907/t20190724\_505559.html},
most of the scientific observations are conducted around the zenith-pointing. 
Thus a zenith attitude of EP and an uniform distribution of the albedo gamma rays 
from the surface of the Earth are assumed for the simulation.

\section{Geant4 mass modeling}
\label{sec::g4}

The mass model of EP/FXT is built under the the framework of 
Geant4 Version 10.5.p1. In this section the detector construction, 
physics process, and generation of the incident particles are introduced.

\subsection{Detector construction}

\begin{figure}[H]
\centering
\includegraphics[width=0.9\columnwidth]{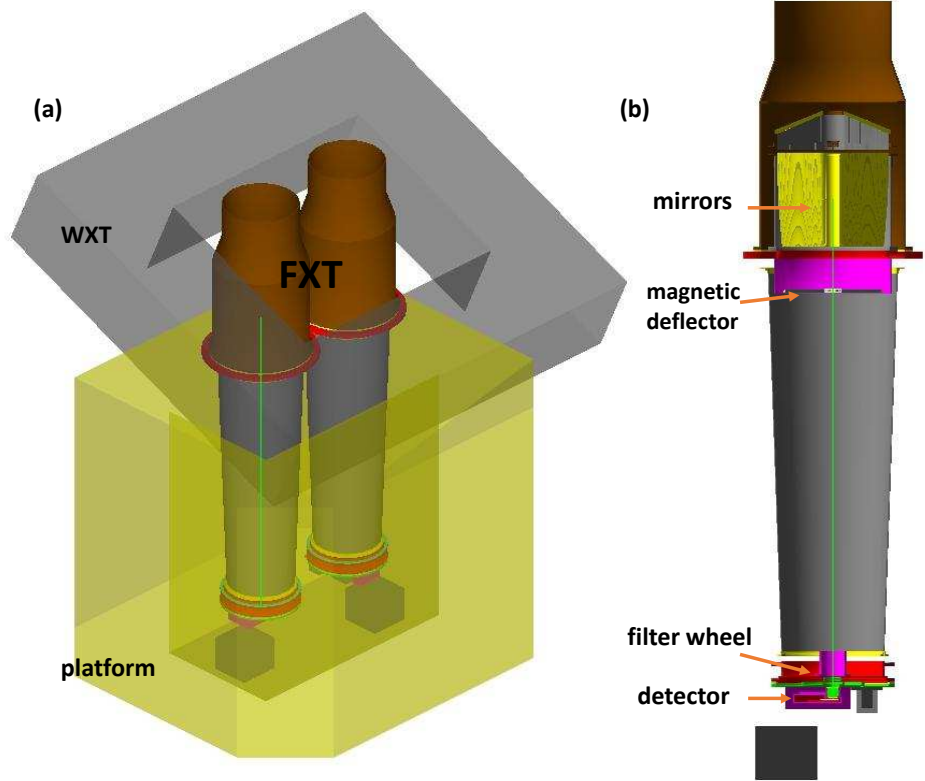}
\caption{(a) The whole structure of EP constructed in the mass model of Geant4.  
(b) The cut-away view of the detailed FXT structure.}
\label{fig::epfxtg4}
\end{figure}

Figure\,\ref{fig::epfxtg4} shows the constructed structure of EP in Geant4. 
Figure\,\ref{fig::epfxtg4}\,(a) illustrates the whole structure of EP. 
It consists of three parts, the platform, the WXT and the FXT. Most of the 
structure is constructed using the Constructive Solid Geometry (CSG) representations of Geant4. 
The platform is shown as the transparent yellow polygons at the bottom and it is built according 
to the realistic geometry dimensions and is filled by the aluminum with an equivalent density that 
is derived from its realistic mass. WXT is simplified as a hollow square around the two FXT modules. 
It is also assigned by the aluminum material with an equivalent density 
derived from its own weight. 
The FXT is constructed in more detail, as shown in Figure\,\ref{fig::epfxtg4}\,(b).
The 54 layers of Wolter-I mirror are built by using the paraboloid of Geant4 
and the extension of hyperboloid developed by Qi et al. \cite{2020NIMPA.96363702Q}. 
The imaging area of the pn-CCD is defined as the sensitive detector, and the 
deposited energy, position and time of each hit on this sensitive detector 
is output for further analysis.

\subsection{Physics}
Given the energy range of interest of FXT and the energies and types of 
the incident particles in the EP orbit, the electromagnetic physics constructor G4EmStandardPhysics\_option4 is chosen, which has an 
optimal mixture for the interaction precision. 
The options of fluorescence (Fluo),
particle induced X-ray emission (PIXE) and auger processes are all set 
to be active. 
For the interaction of photons incident with grazing angles on the Wolter-I mirror, 
the physics model developed by Qi et al. \cite{2020NIMPA.96363702Q} is used 
with a micro-roughness of 0.5\,nm.
And for the scattering of low-energy protons in the mirror, the single
scattering model is used. For silicon detectors,  the 
radioactive background is generally not important and thus this process is 
not included in current simulation.
These are detailed physical processes involved in the silicon-based detectors \cite{2013ITNS...60.3150G}, 
e.g. the electron-hole pair production, 
the charge splitting and charge collections, which are not included in the simulation currently.

\subsection{Primary Generation}
\begin{figure}[H]
\centering
\includegraphics[width=0.6\columnwidth]{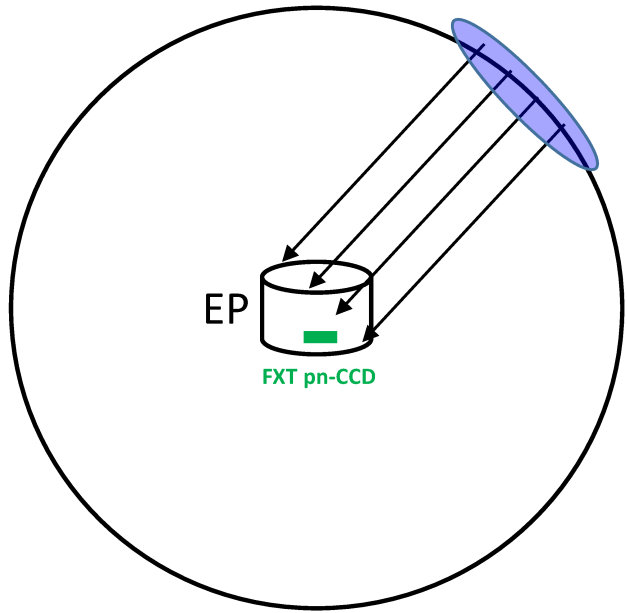}
\caption{The direction and position generation in Geant4 mass modeling of EP/FXT.}
\label{fig::genprimary}
\end{figure}

Generally, the particle type, time, energy, position and direction 
are needed as the input parameters of the 
Geant4 simulation. The energy is easily obtained according to the spectral 
distribution of each background type as presented in Section\,\ref{sec::spcenv}. 
While considering the large structure of 
the EP satellite and the small sensitive geometry of FXT, 28.8mm$\times$28.8\,mm, 
to improve the simulation efficiency, a  method for sampling the position 
and direction of incident particles is introduced.
Figure\,\ref{fig::genprimary} illustrates how the direction and position are sampled.
Firstly, a circle with a diameter large enough to cover 
the cross section of EP is chosen, shown as the blue plane in Figure\,\ref{fig::genprimary}.
Then a direction is chosen according to the direction distribution of each particle type. 
Finally the position is generated by putting this circle plane in that direction, and sampling uniformly
on the circle plane.

\section{Simulation Results}
\label{sec::result}

In this section, the background and sensitivity results of FXT 
are given. The hit information, i.e. the deposited energy, location, 
and time of each hit on the imaging area, output from Geant4 
simulation are analyzed to obtain the background of FXT. Almost 
all hitting time of the responses of one incident event 
on the FPD are within the frame time of the pn-CCD. 
Though the charge splitting mode is not included in the simulation process, 
incident particles might also cause responses on more than one pixel 
(as illustrated in Figure\,\ref{fig::primaryP} 
for the tracks caused by the cosmic ray protons). 
These background events are easily to be discarded. So the first step 
of the data analysis is to assign the hit information to the 
384$\times$384 pixels of the pn-CCD. 
The background events that deposit energy on more than three pixels 
are discarded.
For the remaining events, their energies are obtained by adding up 
the deposited energy on all the hitting pixels. 
The positions of the background events are calculated by 
averaging the pixel center locations weighted by their deposited energies.

\subsection{Background}

The background contributions to the FPD of FXT are classified 
into two types, the FOV background and the instrumental background. 
The former one refers to the background caused by the focusing of the mirrors, 
such as the reflected cosmic photon background and the funneled low energy protons. 
The latter is induced by space particles that interact with the instrument and materials 
around the FPD and finally hit on the pn-CCD, e.g. the background induced by 
cosmic ray protons and electrons, by higher energy photons of the cosmic photon background and 
by albedo gamma rays, etc.

\begin{figure*}
\centering
\includegraphics[width=\linewidth]{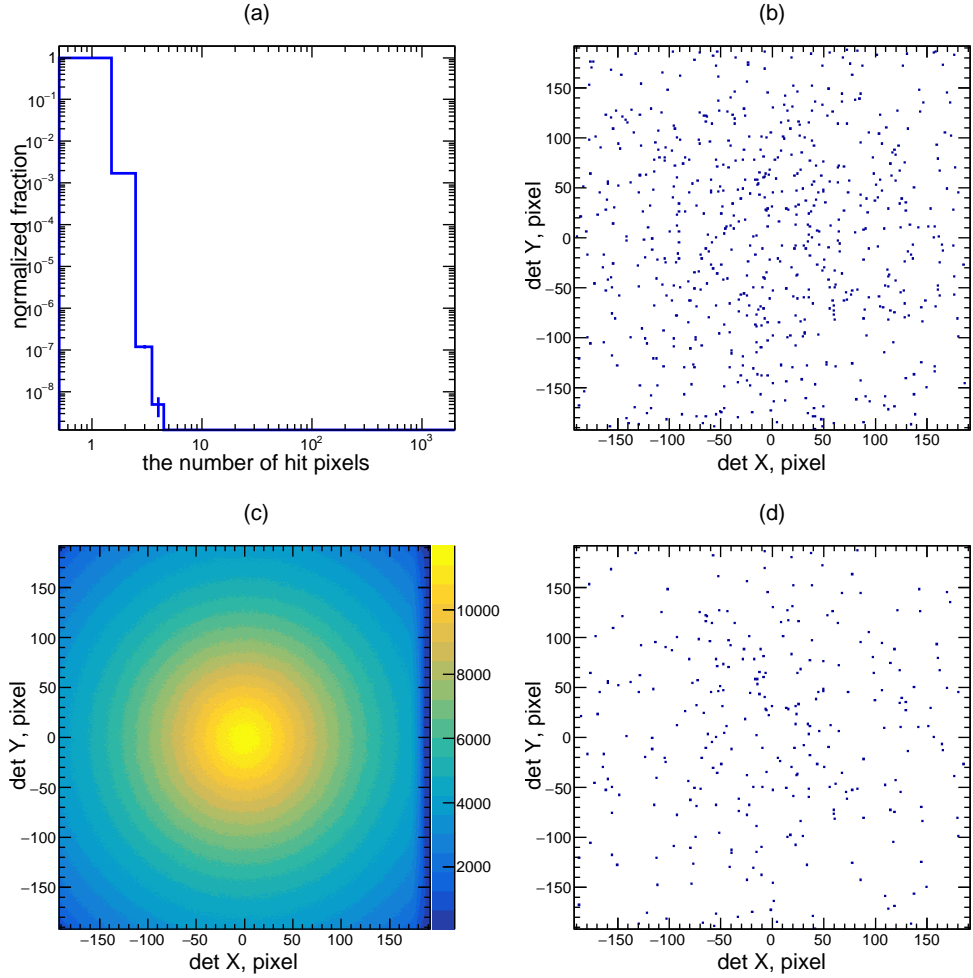}
\caption{The characteristic of FOV background induced by cosmic photons.
(a) The statistical distribution of the number of pixels that one background event
deposits energy on; (b) Tracks of the background events without energy and pixel screening 
for an exposure of $\sim$201 seconds; 
(c) The background event distribution on the imaging area after data screening and this distribution presents the vignetting effect of the Wolter-I mirror; 
(d) The same as panel (b) for background events after an energy (0.5-10 keV) 
and pixel (deposited pixel number$<$4) screening. }
\label{fig::cxbwfi_dis}
\end{figure*}

\begin{figure*}
\centering
\includegraphics[width=\linewidth]{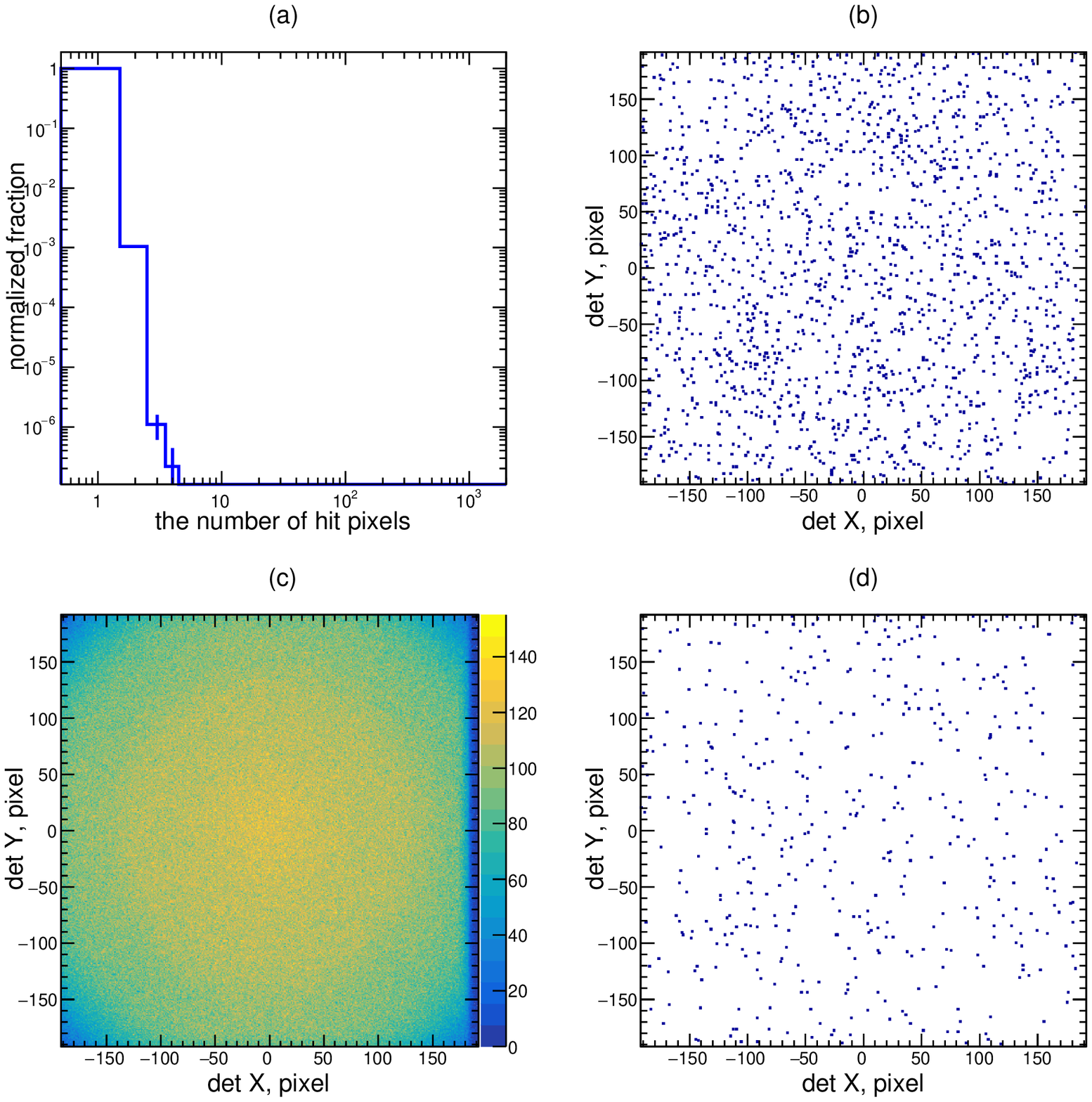}
\caption{The characteristic of FOV background induced by equatorial low-energy
protons. Each panel gives the same result as Figure\,\ref{fig::cxbwfi_dis}, except 
that (b) is the tracks about 212 seconds.}
\label{fig::lowP}
\end{figure*}

For the FOV background of cosmic photons, Figure\,\ref{fig::cxbwfi_dis} 
shows the distribution of the number of hit pixels, 
the tracks of some background events and the background 
event distribution on the FPD. 
The same plots for the background induced by equatorial low energy protons are also 
presented in Figure\,\ref{fig::lowP}.
It can be seen that all the FOV background events hit no more than four pixels and 
most of them hit only one pixel. 
Due to the focusing effect of the Wolter-I mirror, 
the distribution of the FOV background on the FPD are not uniform, and 
more concentrated towards the center of the FOV.

\begin{figure*}
\centering
\includegraphics[width=\linewidth]{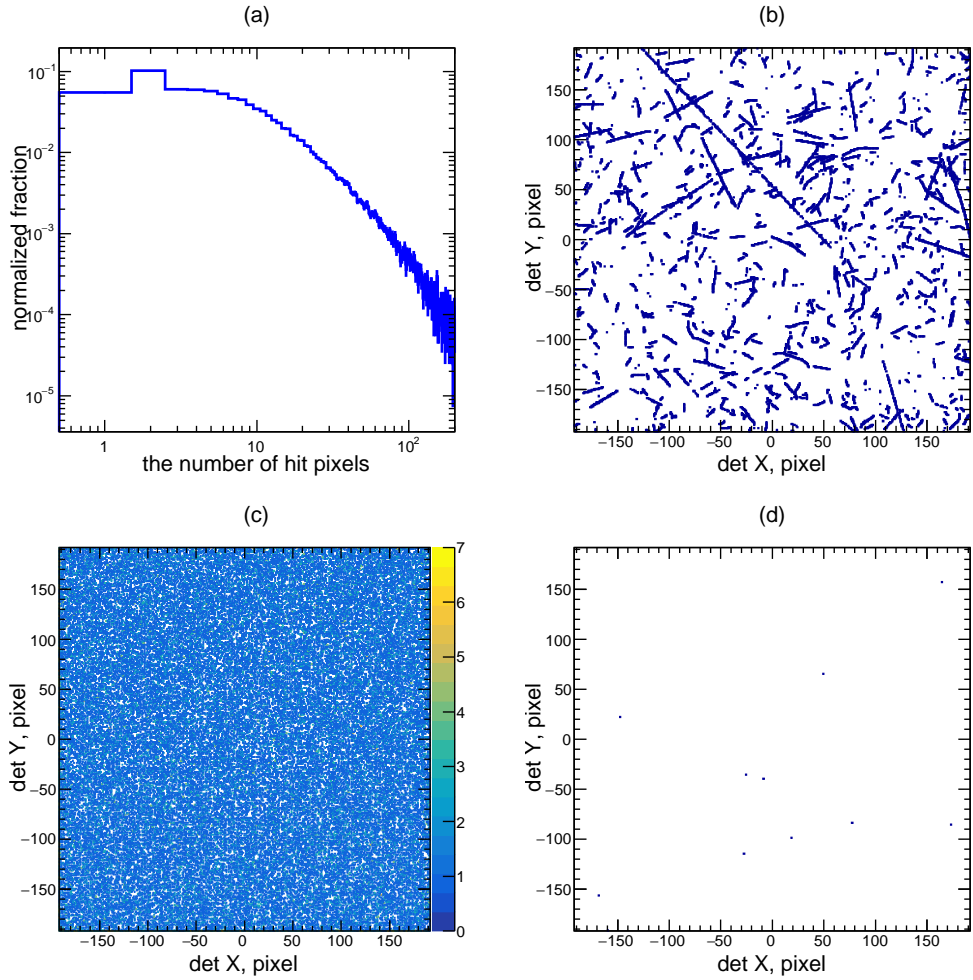}
\caption{The characteristic of instrumental background induced by 
primary cosmic ray protons. 
Each panel plots the same result as Figure\,\ref{fig::cxbwfi_dis}, except 
that (b) is the tracks about 122 seconds.}
\label{fig::primaryP}
\end{figure*}

\begin{figure*}
\centering
\includegraphics[width=\linewidth]{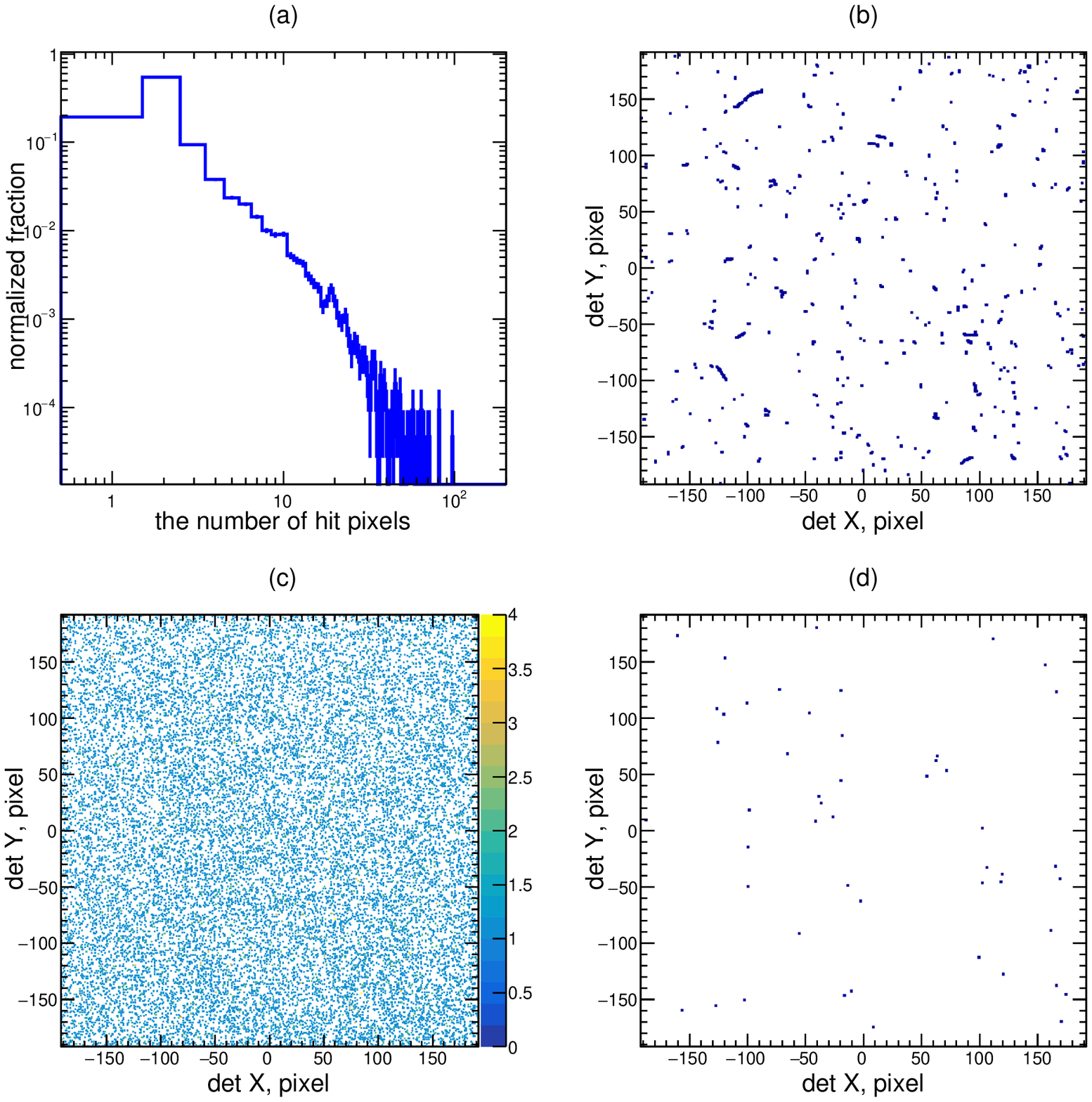}
\caption{The characteristic of instrumental background induced by cosmic photons. 
Each panel plots the same result as Figure\,\ref{fig::cxbwfi_dis}, except 
that (b) is the tracks about 395 seconds.}
\label{fig::hecxb_dis}
\end{figure*}

For the instrumental background, more pixels are impacted, and their distribution on the FPD 
is uniform, as shown in Figure\,\ref{fig::primaryP} and Figure\,\ref{fig::hecxb_dis} 
for the instrumental background caused by primary cosmic ray protons and cosmic photons, 
respectively.
It is worth noting that the instrumental backgrounds are obtained by setting 
the filter wheel to the closed position in the Geant4 mass model.
It can be seen from these figures that long tracks could be 
caused especially by the cosmic ray protons. 
About 16\% background events of cosmic photons hit more that three pixels, while 
this value is 78\% for primary cosmic ray protons.

\begin{figure}[H]
\centering
\includegraphics[width=0.99\columnwidth]{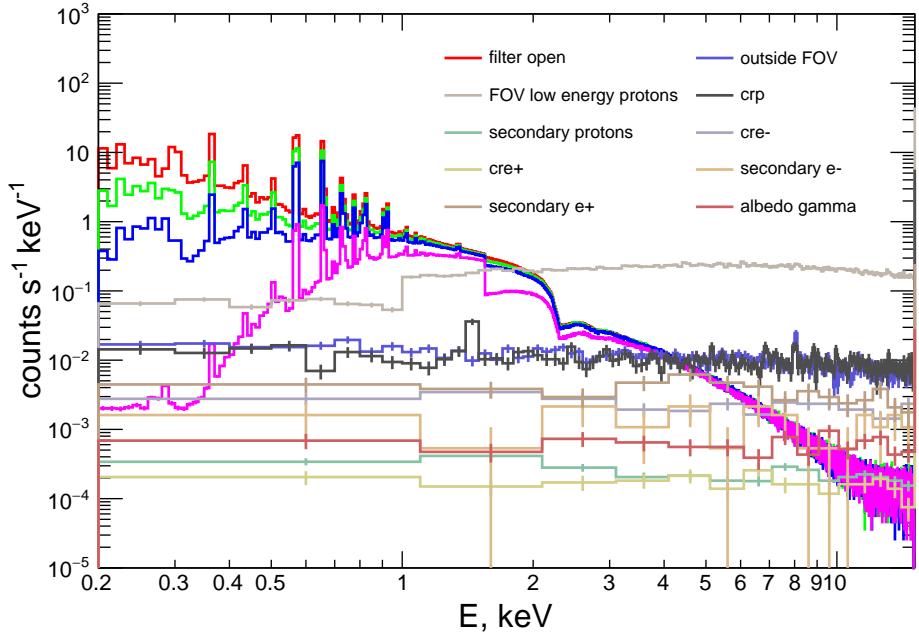}
\caption{The background spectra caused by various space components 
on the imaging area of the FPD of FXT.
The red (green, blue and magenta) line denotes the FOV background of cosmic photons  
corresponding to the open (thin, medium and thick) filter position 
on the filter wheel.
The ``FOV low energy protons" represents the background induced by the low energy protons near the geomagnetic equator through the funneling effect of the mirrors. 
For the  instrumental background components, the ``outside FOV" shows that induced by 
cosmic photons, ``crp" by primary cosmic ray protons, ``secondary protons" by 
secondary cosmic ray protons, ``cre-" by primary cosmic ray electrons, 
``cre+" by primary cosmic ray positrons, ``secondary e-" by secondary 
cosmic ray electrons, ``secondary e+" by secondary cosmic positrons and 
``albedo gamma" by albedo gamma rays. 
}
\label{fig::all_bkg}
\end{figure}

Figure\,\ref{fig::all_bkg} gives the background spectra of various components 
on the imaging area of the FPD of FXT after getting rid of events that hit more than three pixels. 
The FOV background of the equatorial low energy protons 
(FOV low energy protons) are given 
corresponding to the open position of the filter wheel.
It can be seen that the instantaneous background level 
of the equatorial low energy protons, $\sim$0.1 counts\,s$^{-1}$\,keV$^{-1}$, 
is much higher than the instrumental background. However,   
these low energy protons only exist near the geomagnetic equator. 
On the one hand, the time intervals when these background events exist 
are easy to be thrown away through the selection of housekeeping data or 
the selection of the background rate above several keV. 
On the other hand, they are easily to be shielded by the materials of 
the filtered positions of the filter wheel. 
Therefore in the following text, the background and sensitivity results 
are all obtained by neglecting this background component.

The instrumental background spectra of primary cosmic ray protons (crp), 
primary cosmic ray electrons (cre-) and positrons (cre+), 
secondary cosmic ray protons (secondary protons), secondary cosmic ray 
electrons (secondary e-) and positrons (secondary e+), 
cosmic photon background (outside FOV), and albedo gamma rays (albedo gamma)
have an almost flat shape among 0.2-12\,keV. 
The most dominant instrumental background components are induced 
by primary cosmic ray protons and cosmic photon background. 
The Cu K$\alpha$ line at $\sim$8\,keV is obvious, which comes from 
the fluorescence line of the copper shielding box around the FPD. 
The line between 1.4-1.5\,keV on the spectrum of cosmic ray protons 
comes from the aluminum of the closed position 
of the filter wheel \cite{2021A&A...647A...1P}.
The total instrumental background in the energy range from 0.5 to 10\,keV 
on the whole imaging area of the FPD corresponds 
to a level of $\sim$3.1$\times$10$^{-2}$ counts\,s$^{-1}$\,keV$^{-1}$, 
i.e. 3.7$\times$10$^{-3}$ counts\,s$^{-1}$\,keV$^{-1}$\,cm$^{-2}$ after 
area normalization.

\begin{figure}[H]
\centering 
\includegraphics[width=0.9\columnwidth]{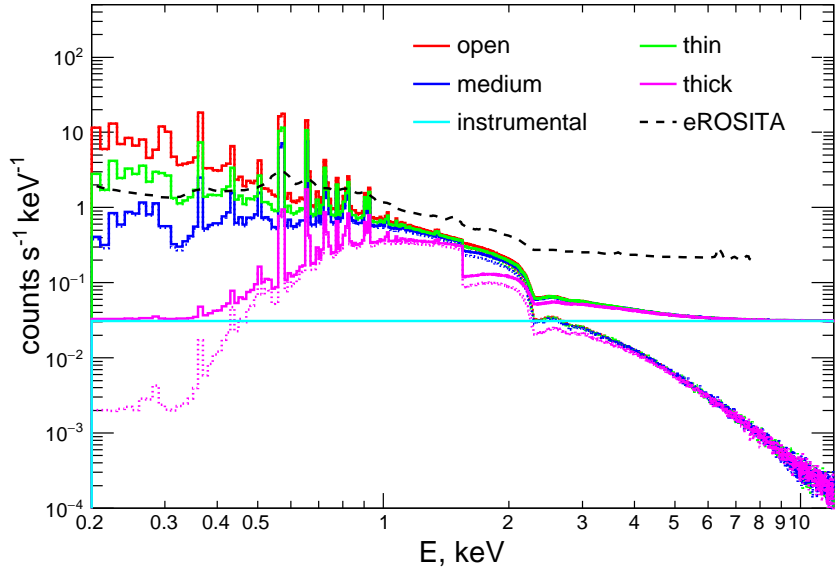}
\caption{The total background spectra on the imaging area of FXT.
The total instrumental background is denoted by the cyan line. 
The solid (dashed) lines are the total background spectra 
(FOV background spectra induced by cosmic photons) for  
different filter wheel positions.
The dashed line in black shows 
the in-orbit background spectrum of one telescope of eROSITA \cite{2021A&A...647A...1P}.
}
\label{fig::compbkg}
\end{figure}

The FOV background spectra of cosmic photons corresponding to 
the open position of the filter wheel (filter open), 
the thin filter position (the green line), the medium filter position 
(the blue line) and the thick filter position (the magenta line) 
are all present in Figure\,\ref{fig::all_bkg}. 
Table\,\ref{tab::bkgrate} lists the FOV and total instrumental 
background rates on the imaging area of the FPD in different energy bands. 
The total background spectra for different filter positions are shown in
Figure\,\ref{fig::compbkg}. The instrumental background spectrum is 
obtained by averaging the sum rates of various instrumental background 
components between 0.5 to 10 keV.
It can be seed that the FOV background dominates below $\sim$2\,keV. 
Above this energy, the instrumental background is overwhelming.
Also plotted in Figure\,\ref{fig::compbkg} is the observed background 
spectrum of one eROSITA telescope \cite{2021A&A...647A...1P}. 
For a single telescope, the FOV background of EP/FXT and eROSITA is comparable, 
while the instrumental background of EP/FXT is about 7 times smaller 
than that of eROSITA. 

\begin{table}[H]
\caption{The background rates of FOV and instrumental background 
on the imaging area of the FPD and within the focal spot of FXT.
The focal spot is taken as a circle with a radius of $30^{\prime\prime}$ 
from the center of the optical focus. Unit: counts\,s$^{-1}$}
\label{tab::bkgrate}
\centering
\begin{threeparttable}
\begin{tabular}{cccc}
\hline
on the FPD   & 0.5-10\,keV & 0.5-2\,keV & 2-10\,keV \\
instrument &  0.29 &   0.06  &   0.23 \\
 open &       1.61   &  1.51   &  0.09   \\
 thin &       1.27   &  1.18   &  0.09   \\
 medium&      1.00   &  0.92  &   0.08  \\
 thick&       0.46  &   0.39  &    0.07 \\
\hline
within focal spot & 0.5-10\,keV & 0.5-2\,keV & 2-10\,keV \\
instrument &  5.6$\times$10$^{-5}$ &   1.2$\times$10$^{-5}$  & 4.4$\times$10$^{-5}$ \\
 open & 7.6$\times$10$^{-4}$ &  7.0$\times$10$^{-4}$  &  5.9$\times$10$^{-5}$ \\
 thin &   6.1$\times$10$^{-4}$ &  5.5$\times$10$^{-4}$  &  6.1$\times$10$^{-5}$ \\
 medium&    4.7$\times$10$^{-4}$ &  4.2$\times$10$^{-4}$  &  5.0$\times$10$^{-5}$ \\
 thick&    2.2$\times$10$^{-4}$ &  1.8$\times$10$^{-4}$  &  4.0$\times$10$^{-5}$ \\
\hline
\end{tabular}
\end{threeparttable}
\end{table}

The general way for an imaging telescope to extract the background 
to be subtracted from the source signal is to choose 
a region (the background region) around the source (the source region) 
on the imaging map. 
Considering the designed angular resolution of FXT,  
a circle with a radius of $30^{\prime\prime}$ is chosen as the source 
region for the on-axis point sources. This region is also used as the focal 
spot region in this work.
Considering the non uniform distribution of the FOV background on the FPD,  
the background spectrum extracted from the background region has to 
be corrected when it is subtracted as the background of the source region.
To avoid the systematic error induced by this correction, 
the background from the source region is extracted 
to give an estimate of the background level within the focal spot of FXT directly. 
The FOV and instrumental background rates %on the whole FPD and 
within the focal spot at different energy bands are listed in Table\,\ref{tab::bkgrate} as well, where
the instrumental background  
is obtained by scaling the instrumental background on the imaging area of the FPD 
to the focal spot region due to its uniform distribution on the imaging area, and the 
FOV background is obtained by counting the events within the focal spot.
In the energy band of 0.5-10\,keV the instrumental background 
within the focal spot corresponds to 5.9$\times$10$^{-6}$ counts\,s$^{-1}$\,keV$^{-1}$.

\subsection{Sensitivity}
\begin{figure}[H]
\centering
\includegraphics[width=0.9\columnwidth]{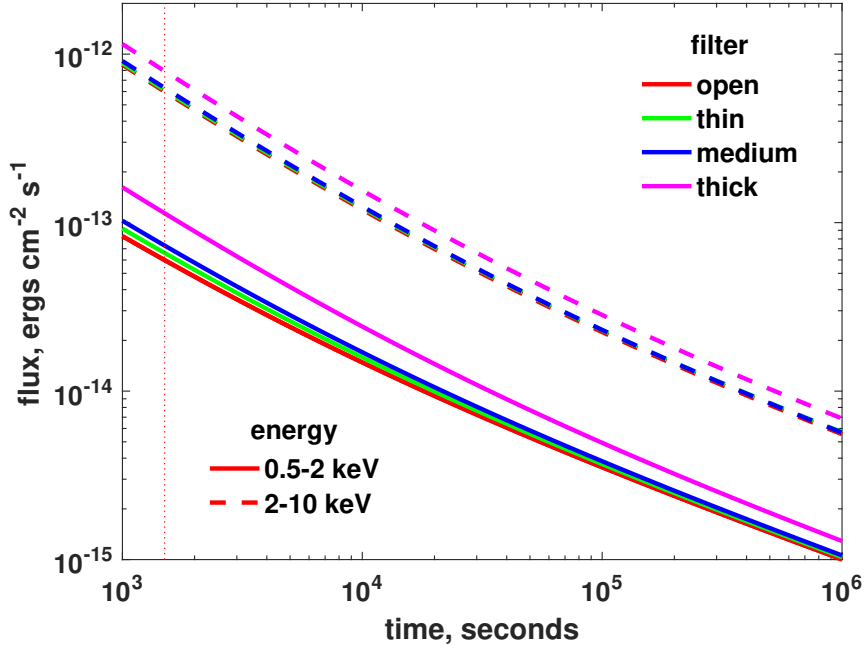}
\caption{The sensitivity of FXT derived from
the ideal case where no background uncertainty is considered
\cite{2018ApJS..236...17V}. Different line types correspond to 
different energy bands (
solid: 0.5-2 keV; dashed: 2-10 keV).
%solid: 0.5-10 keV; dotted: 0.5-2 keV; dashed: 2-10 keV).
Different line colors denote different filter 
wheel positions (red: open; green: thin; blue: medium; magenta: thick).
Sensitivity is computed for an assumed Crab spectrum, i.e. 
an absorbed power-law spectrum with an index of 2.05 and 
a column density N$_{\rm H}$=2$\times$10$^{21}$\,cm$^{-2}$.
The vertical dotted line marks the typical exposure time during EP survey, 
i.e. 1500 seconds.
}
\label{fig::ideal_sensitivity}
\end{figure}

The sensitivity of EP/FXT could be estimated based on the simulated background.
Here the sensitivity is defined as the flux limit of a weak source which can be 
detected at the $5\sigma$ significance level.
There are several ways to calculate the detection significance of a source in high energy astrophysics measurements, 
for example the commonly used Li\&Ma method \cite{1983ApJ...272..317L} and the 
specific methods taking into account the impact of different kinds of 
background uncertainties 
\cite{2018ApJS..236...17V}. In this work, the ideal case \cite{2018ApJS..236...17V} where 
no background uncertainty is adopted. This corresponds to 
the best theoretical sensitivity limit. And the likelihood ratio method of 
Li\&Ma method \cite{1983ApJ...272..317L} are also utilized to have a comparison. 
For the ideal case, the expected counts from the source is 
\cite{2018ApJS..236...17V}
\begin{equation}
\label{eq::SNR1}
M = 11.090 + 7.415\times\sqrt{B} 
\end{equation}
for the detection efficiency of 99\% at 5$\sigma$ level, 
where M is the source counts and B the background counts. 
Based on the simulated background rate within the focal spot,
the sensitivity of FXT is given in Figure\,\ref{fig::ideal_sensitivity}.
It can be seen that, by assuming a Crab-like source spectrum, for once typical pointing observation during EP survey, 
i.e. an exposure of 25 \,minutes,  
the sensitivity of FXT for different filter setup could achieve a level of $(5.9\sim11.3)\times10^{-14}$ erg cm$^{-2}$ s$^{-1}$ (5.1$\sim$9.7\,$\mu$crab)
in the soft energy band 0.5-2 keV and $(5.9\sim7.9)\times10^{-13}$ erg cm$^{-2}$ s$^{-1}$ (28.0$\sim$37.1\,$\mu$crab) in the hard energy band 2-10 keV.

For the likelihood ratio method, the significance \cite{1983ApJ...272..317L}
\begin{equation}
\label{eq::lima}
S = \sqrt{ 2 \times \left\{ 
M \ln \left[ \frac{1+\alpha}{\alpha}\left(\frac{M}{M+B}\right) \right]  + 
B \ln \left[ (1+\alpha)\left(\frac{B}{M+B}\right) \right] 
\right\} }.
\end{equation}
For FXT, $\alpha$ is the ratio of the backgrounds between the source 
region and the background region. 
For simplicity, we have taken the background from the focal spot, 
therefore $\alpha=1$. 
The calculated flux for a source with the spectral shape 
the same as Crab to be detected with an exposure of 25 minutes is $(1.2\sim2.2)\times10^{-13}$ erg cm$^{-2}$ s$^{-1}$ (10.6$\sim$18.7\,$\mu$crab)
in the soft energy band (0.5-2 keV) for different filter positions and $(10.3\sim13.5)\times10^{-13}$ erg cm$^{-2}$ s$^{-1}$ (48.5$\sim$64\,$\mu$crab) in the hard energy band (2-10 keV).
These values are a bit larger than those shown in 
Figure\,\ref{fig::ideal_sensitivity}
%Table\,\ref{tab::SNR1}, 
because the Poisson distribution of the background is included in this method.
In addition, given the uneven background distribution 
of the cosmic photons on the FPD of FXT, 
more uncertainties will be introduced and the significance will be reduced
in the real observation and data analysis procedure
if background counts come from model methods. 
Then higher fluxes are needed to achieve a 5$\sigma$ threshold. 
For example, if a 10\% systematic uncertainty on the background estimate is considered, the obtained sensitivity for an exposure of 25 minutes is 
$(1.6\sim2.6)\times10^{-13}$ erg cm$^{-2}$ s$^{-1}$ (13.6$\sim$19.9\,$\mu$crab)
in the soft energy band (0.5-2 keV) for different filter positions and $(11.0\sim14.5)\times10^{-13}$ erg cm$^{-2}$ s$^{-1}$ (52$\sim$69\,$\mu$crab) in the hard energy band (2-10 keV).

\begin{figure}[H]
\centering
\includegraphics[width=0.9\columnwidth]{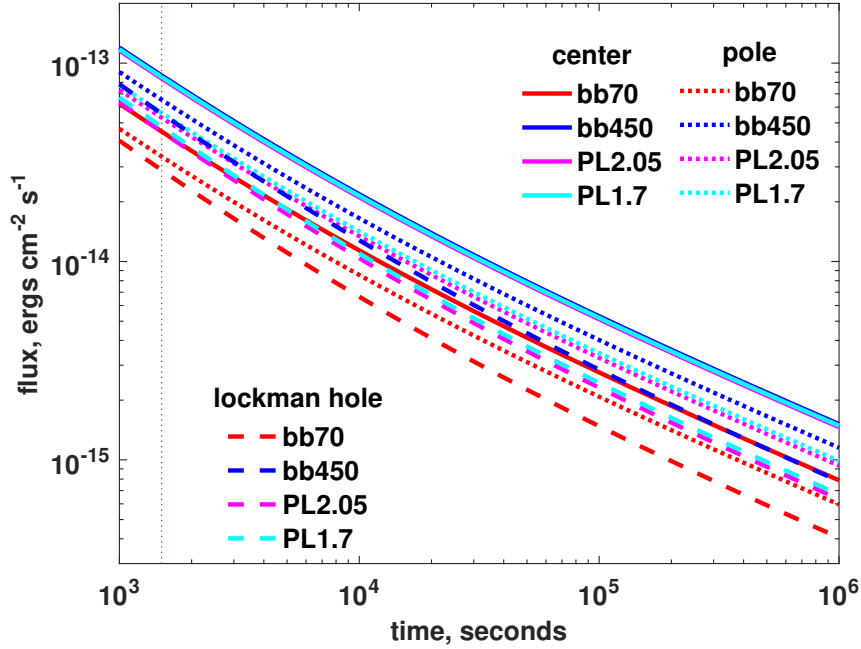}
\caption{The sensitivity of FXT for different pointing directions and 
source spectra derived from the ideal case. All the sensitivities plotted here are obtained 
for the energy band of 0.5-2\,keV when the filter wheel is set to open position. 
Solid lines correspond to the Galactic center pointing direction with 
a column density N$_{\rm H}\sim10^{22}$\,cm$^{-2}$, 
dotted lines Galactic pole with N$_{\rm H}\sim10^{20}$\,cm$^{-2}$, 
and dashed lines the lockman hole with N$_{\rm H}\sim5\times10^{19}$\,cm$^{-2}$.
Red lines indicate the black body spectrum with a temperature of 70\,eV, 
blue lines the black body spectrum with a temperature of 450\,eV, 
magenta lines the power-law spectrum with an index of 2.05 and cyan lines 
the power-law spectrum with an index of 1.7.
The vertical dotted line marks an exposure of 1500 seconds.
}
\label{fig::dir_spc_sensitivity}
\end{figure}

The sensitivity also depends on the sky area and the spectral characteristics of
the source \cite{xmmuhb}. The Galactic Soft X-ray Background 
and the Galactic hydrogen column density (N$_{\rm H}$) 
vary with the observational pointing direction. 
To investigate the impact on the sensitivity, 
we obtained the Soft X-ray Background 
at the direction of Galactic center, Galactic pole and the lockman hole
using the X-ray Background Tool from 
the HEASARC website\footnote{https://heasarc.gsfc.nasa.gov/cgi-bin/Tools/xraybg/xraybg.pl}, 
and also got Galactic N$_{\rm H}$ there. 
Compared with the soft X-ray diffuse background we used in the simulation, 
the Soft X-ray Background at 0.5-2 keV needs to be scale by a factor of 1.18, 1.22 and 0.53, 
respectively for the Galactic center, Galactic pole and the lockman hole direction.
For the intrinsic spectrum of the target source, 
two black body spectra with temperatures of 70\,eV and 450\,eV, respectively 
and two power-law spectra with indices of 1.7 and 2.05, respectively 
were considered. 
The obtained sensitivities in the energy band of 0.5-2\,keV 
are plotted in Figure\,\ref{fig::dir_spc_sensitivity} 
for these three directions and four spectral shapes. 
The best result is obtained at the lockman hole direction. 
The sensitivity with an exposure of 25 minutes varies from 
2.9$\times10^{-14}$ erg cm$^{-2}$ s$^{-1}$ 
to 8.7$\times10^{-14}$ erg cm$^{-2}$ s$^{-1}$. 
Compared with the point source sensitivity of the 7 telescopes of eROSITA \cite{2021A&A...647A...1P}, which is 1.1$\times10^{-14}$ erg cm$^{-2}$ s$^{-1}$ in  0.2-2.3\,keV, the estimated sensitivity of one FXT module is reasonable in this work.

\section{Conclusions}
\label{conclusion}

In this work, we have obtained an estimate of the in-orbit background of EP/FXT 
based on simulations using the Geant4 toolkit. We have also given an estimate of its sensitivity 
based on the obtained background level. 
During simulation, the energy deposition on the imaging area 
of the FPD is recorded. 
For the space radiation environment, 
the diffuse photon background including the extra-galactic background and 
the Galactic foreground, the cosmic ray protons, electrons and 
positrons, as well as the albedo gamma rays and the low-energy protons 
near the magnetic equator are all considered. 
Events that deposited energies on more than three pixels are discarded. 
The background induced by the equatorial low-energy protons is about one 
order of magnitude higher than the instrumental background, but it is easily 
shielded or removed during data analysis and processing. 
The estimated background level and 
sensitivity presented in this paper are obtained without the contribution of 
this component.
The instrumental background distributes uniformly on the FPD, and 
its background level is about 
3.1$\times$10$^{-2}$ counts s$^{-1}$\,keV$^{-1}$ 
(3.7$\times$10$^{-3}$ counts s$^{-1}$\,keV$^{-1}$\,cm$^{-2}$ )
on the whole imaging area of the FPD and  
5.6$\times$10$^{-5}$ counts s$^{-1}$\,keV$^{-1}$
within the focal spot with a radius of 30$^{\prime\prime}$.
The FOV background caused by cosmic photons dominates below $\sim$2\,keV, 
and the distribution is more concentrated towards the FOV center.
Its background rate between 0.5-10\,keV is 1.61 counts s$^{-1}$ on the whole FPD and 
7.6$\times$10$^{-4}$ counts s$^{-1}$ within the focal spot when the filter wheel 
is set to the open position. While this rate reduces to 0.46 counts s$^{-1}$ on the whole FPD 
and 2.2$\times$10$^{-4}$ counts s$^{-1}$ within the focal spot 
when the filter wheel is set to the thick position.
Based on the background level within the focal spot and 
assuming a Crab-like source spectrum, 
the estimated ideal sensitivity of FXT could achieve  
$(5.9\sim11.3)\times10^{-14}$ erg cm$^{-2}$ s$^{-1}$ 
for an exposure of 25\,minutes in the energy band of 0.5-2 keV, 
and $(5.9\sim7.9)\times10^{-14}$ erg cm$^{-2}$ s$^{-1}$ 
in the energy band of 2-10 keV, 
when the filter wheel is set from open to thick position.
If the uncertainty from Poisson distribution is considered, 
the sensitivity becomes 
$(1.2\sim2.2)\times10^{-13}$ erg cm$^{-2}$ s$^{-1}$ in 0.5-2 keV and 
$(10.3\sim13.5)\times10^{-13}$ erg cm$^{-2}$ s$^{-1}$ in 2-10 keV.
Furthermore, if an additional 10\% uncertainty from the background 
subtraction is included, the sensitivity becomes  
$(1.6\sim2.6)\times10^{-13}$ erg cm$^{-2}$ s$^{-1}$ in 0.5-2 keV and 
$(11.0\sim14.5)\times10^{-13}$ erg cm$^{-2}$ s$^{-1}$ in 2-10 keV, 
about 2 times worse than the idea case.

For the estimated background, it is worth noting that the result 
given in this work is obtained by adopting the averaged 
space environment radiation models and zenith satellite pointing attitude. 
The instrumental background induced by cosmic rays could vary by a factor of $\sim$3
from low geomagnetic latitudes to high latitudes
according to the background observation of Insight-HXMT \cite{2020JHEAp..27...24L}. 
The response of FXT, e.g. the ancillary response file, 
the point spread function and the vignetting function, 
are not calibrated 
due to the lack of ground experimental data currently. Therefore, 
the FOV background might sort of differ from the real measurement after the launch, and it also varies with the pointing direction in the sky. 
The soft X-ray background below 2 keV varies by a factor of $\sim2$ from 
the lockman hole direction to the Galactic center direction according 
to the ROSAT All-Sky Survey diffuse background maps. 
The sensitivity not only changes with the background and  
uncertainty, but also with the pointing direction and the source spectrum. 
Beside the Crab-like spectrum of a power-law shape with an index of 2.05, 
we also estimate the sensitivity by assuming a softer power-law spectrum with 
the index of 1.7 and two black-body spectra with the temperature of 70 eV 
and 450 eV, respectively.  
The obtained theoretical sensitivity varies from 
$2.9\times10^{-14}$ erg cm$^{-2}$ s$^{-1}$ 
to $8.7\times10^{-14}$ erg cm$^{-2}$ s$^{-1}$  
in the energy band of 0.5-2 keV 
for these four spectra at different pointing directions
when the open filter wheel position is set.
The estimated in-orbit background level and sensitivity of FXT in this 
work will help conduct the observation simulation, 
make up the observation strategy and pre-study the scientific targets.

\section*{Declaration of Competing Interest}
The authors declare that they have no known competing financial interests or personal 
relationships that could have appeared to influence the work reported in this paper.

\section*{Acknowledgements}
We are grateful to all the colleagues in EP team. 
J. Zhang thank Z. C. Tang for the discussion with AMS-02 data. 
This work is supported by the Einstein-Probe (EP) Program 
which is funded by the Strategic Priority Research Program of the Chinese Academy of 
Sciences Grant No.XDA15310000. 
We also acknowledge Project 12003037 supported by the National Natural Science Foundation of China.


\bibliographystyle{elsarticle-num} 
\bibliography{cas-refs}

\begin{thebibliography}{10}
\expandafter\ifx\csname url\endcsname\relax
  \def\url#1{\texttt{#1}}\fi
\expandafter\ifx\csname urlprefix\endcsname\relax\def\urlprefix{URL }\fi
\expandafter\ifx\csname href\endcsname\relax
  \def\href#1#2{#2} \def\path#1{#1}\fi

\bibitem{2018SPIE10699E..25Y}
W.~{Yuan}, C.~{Zhang}, Z.~{Ling}, et~al., {Einstein Probe: a lobster-eye
  telescope for monitoring the x-ray sky}, in: J.-W.~A. {den Herder},
  S.~{Nikzad}, K.~{Nakazawa} (Eds.), Space Telescopes and Instrumentation 2018:
  Ultraviolet to Gamma Ray, Vol. 10699 of Society of Photo-Optical
  Instrumentation Engineers (SPIE) Conference Series, 2018, p. 1069925.
\newblock \href {https://doi.org/10.1117/12.2313358}
  {\path{doi:10.1117/12.2313358}}.

\bibitem{2016SSRv..202..235Y}
W.~{Yuan}, L.~{Amati}, J.~K. {Cannizzo}, et~al., {Perspectives on Gamma-Ray
  Burst Physics and Cosmology with Next Generation Facilities}, \ssr 202~(1-4)
  (2016) 235--277.
\newblock \href {http://arxiv.org/abs/1606.09536} {\path{arXiv:1606.09536}},
  \href {https://doi.org/10.1007/s11214-016-0274-z}
  {\path{doi:10.1007/s11214-016-0274-z}}.

\bibitem{2018SSPMA..48c9502Y}
W.~{Yuan}, C.~{Zhang}, Y.~{Chen}, et~al., {Einstein Probe: Exploring the
  ever-changing X-ray Universe}, Scientia Sinica Physica, Mechanica \&
  Astronomica 48~(3) (2018) 039502.
\newblock \href {https://doi.org/10.1360/SSPMA2017-00297}
  {\path{doi:10.1360/SSPMA2017-00297}}.

\bibitem{2017ExA....43..267Z}
D.~{Zhao}, C.~{Zhang}, W.~{Yuan}, et~al., {Geant4 simulations of a wide-angle
  x-ray focusing telescope}, Experimental Astronomy 43~(3) (2017) 267--283.
\newblock \href {http://arxiv.org/abs/1703.09380} {\path{arXiv:1703.09380}},
  \href {https://doi.org/10.1007/s10686-017-9534-5}
  {\path{doi:10.1007/s10686-017-9534-5}}.

\bibitem{2010SPIE.7742E..0YT}
C.~{Tenzer}, G.~{Warth}, E.~{Kendziorra}, A.~{Santangelo}, {Geant4 simulation
  studies of the eROSITA detector background}, in: A.~D. {Holland}, D.~A.
  {Dorn} (Eds.), High Energy, Optical, and Infrared Detectors for Astronomy IV,
  Vol. 7742 of Society of Photo-Optical Instrumentation Engineers (SPIE)
  Conference Series, 2010, p. 77420Y.
\newblock \href {https://doi.org/10.1117/12.857087}
  {\path{doi:10.1117/12.857087}}.

\bibitem{2012ExA....33...39P}
E.~{Perinati}, C.~{Tenzer}, A.~{Santangelo}, et~al., {The radiation environment
  in L-2 orbit: implications on the non-X-ray background of the eROSITA pn-CCD
  cameras}, Experimental Astronomy 33~(1) (2012) 39--53.
\newblock \href {https://doi.org/10.1007/s10686-011-9269-7}
  {\path{doi:10.1007/s10686-011-9269-7}}.

\bibitem{Weidenspointner:2008oka}
G.~Weidenspointner, M.~G. Pia, A.~Zoglauer, {Application of the Geant4 PIXE
  implementation for space missions new models for PIXE simulation with
  Geant4}, in: {2008 IEEE Nuclear Science Symposium and Medical Imaging
  Conference and 16th International Workshop on Room-Temperature Semiconductor
  X-Ray and Gamma-Ray Detectors}, 2008, p. 2877--2884.
\newblock \href {https://doi.org/10.1109/NSSMIC.2008.4774969}
  {\path{doi:10.1109/NSSMIC.2008.4774969}}.

\bibitem{2016SPIE.9905E..6WF}
V.~{Fioretti}, A.~{Bulgarelli}, G.~{Malaguti}, et~al., {Monte Carlo simulations
  of soft proton flares: testing the physics with XMM-Newton}, in: J.-W.~A.
  {den Herder}, T.~{Takahashi}, M.~{Bautz} (Eds.), Space Telescopes and
  Instrumentation 2016: Ultraviolet to Gamma Ray, Vol. 9905 of Society of
  Photo-Optical Instrumentation Engineers (SPIE) Conference Series, 2016, p.
  99056W.
\newblock \href {http://arxiv.org/abs/1607.05319} {\path{arXiv:1607.05319}},
  \href {https://doi.org/10.1117/12.2232537} {\path{doi:10.1117/12.2232537}}.

\bibitem{2013ExA....36..451C}
R.~{Campana}, M.~{Feroci}, E.~{Del Monte}, et~al., {Background simulations for
  the Large Area Detector onboard LOFT}, Experimental Astronomy 36~(3) (2013)
  451--477.
\newblock \href {http://arxiv.org/abs/1305.3789} {\path{arXiv:1305.3789}},
  \href {https://doi.org/10.1007/s10686-013-9341-6}
  {\path{doi:10.1007/s10686-013-9341-6}}.

\bibitem{2018Galax...6...50X}
F.~{Xie}, M.~{Pearce}, {A Study of Background Conditions for
  Sphinx{\textemdash}The Satellite-Borne Gamma-Ray Burst Polarimeter}, Galaxies
  6~(2) (2018) 50.
\newblock \href {http://arxiv.org/abs/1809.06629} {\path{arXiv:1809.06629}},
  \href {https://doi.org/10.3390/galaxies6020050}
  {\path{doi:10.3390/galaxies6020050}}.

\bibitem{2015Ap&SS.360...47X}
F.~{Xie}, J.~{Zhang}, L.-M. {Song}, et~al., {Simulation of the in-flight
  background for HXMT/HE}, \apss 360 (2015) 13.
\newblock \href {http://arxiv.org/abs/1511.02997} {\path{arXiv:1511.02997}},
  \href {https://doi.org/10.1007/s10509-015-2559-1}
  {\path{doi:10.1007/s10509-015-2559-1}}.

\bibitem{2020Ap&SS.365..158Z}
J.~{Zhang}, X.~{Li}, M.~{Ge}, et~al., {Comparison of simulated backgrounds with
  in-orbit observations for HE, ME, and LE onboard Insight-HXMT}, \apss 365~(9)
  (2020) 158.
\newblock \href {http://arxiv.org/abs/2009.14489} {\path{arXiv:2009.14489}},
  \href {https://doi.org/10.1007/s10509-020-03873-8}
  {\path{doi:10.1007/s10509-020-03873-8}}.

\bibitem{Agostinelli:2002hh}
S.~Agostinelli, et~al., {GEANT4--a simulation toolkit}, Nucl. Instrum. Meth. A
  506 (2003) 250--303.
\newblock \href {https://doi.org/10.1016/S0168-9002(03)01368-8}
  {\path{doi:10.1016/S0168-9002(03)01368-8}}.

\bibitem{Allison:2006ve}
J.~Allison, et~al., {Geant4 developments and applications}, IEEE Trans. Nucl.
  Sci. 53 (2006) 270.
\newblock \href {https://doi.org/10.1109/TNS.2006.869826}
  {\path{doi:10.1109/TNS.2006.869826}}.

\bibitem{Allison:2016lfl}
J.~Allison, et~al., {Recent developments in Geant4}, Nucl. Instrum. Meth. A 835
  (2016) 186--225.
\newblock \href {https://doi.org/10.1016/j.nima.2016.06.125}
  {\path{doi:10.1016/j.nima.2016.06.125}}.

\bibitem{2009NIMPA.599..260B}
E.~J. {Buis}, G.~{Vacanti}, {X-ray tracing using Geant4}, Nuclear Instruments
  and Methods in Physics Research A 599~(2-3) (2009) 260--263.
\newblock \href {http://arxiv.org/abs/0810.1273} {\path{arXiv:0810.1273}},
  \href {https://doi.org/10.1016/j.nima.2008.11.002}
  {\path{doi:10.1016/j.nima.2008.11.002}}.

\bibitem{2020NIMPA.96363702Q}
L.~Q. {Qi}, G.~{Li}, Y.~P. {Xu}, et~al., {Geant4 simulation for the responses
  to X-rays and charged particles through the eXTP focusing mirrors}, Nuclear
  Instruments and Methods in Physics Research A 963 (2020) 163702.
\newblock \href {https://doi.org/10.1016/j.nima.2020.163702}
  {\path{doi:10.1016/j.nima.2020.163702}}.

\bibitem{1980JETP...52..225R}
V.~S. {Remizovich}, M.~I. {Ryazanov}, I.~S. {Tilinin}, {Energy and angular
  distributions of particles reflected in glancing incidence of a beam of ions
  on the surface of a material}, Soviet Journal of Experimental and Theoretical
  Physics 52 (1980) 225.

\bibitem{2020ExA....49..115A}
R.~{Amato}, T.~{Mineo}, A.~{D'A{\i}}, et~al., {Soft proton scattering at
  grazing incidence from X-ray mirrors: analysis of experimental data in the
  framework of the non-elastic approximation}, Experimental Astronomy 49~(3)
  (2020) 115--140.
\newblock \href {http://arxiv.org/abs/2003.07295} {\path{arXiv:2003.07295}},
  \href {https://doi.org/10.1007/s10686-020-09657-w}
  {\path{doi:10.1007/s10686-020-09657-w}}.

\bibitem{2017ExA....44..401G}
A.~{Guzm{\'a}n}, E.~{Perinati}, S.~{Diebold}, et~al., {A revision of soft
  proton scattering at grazing incidence and its implementation in the geant4
  toolkit}, Experimental Astronomy 44~(3) (2017) 401--411.
\newblock \href {https://doi.org/10.1007/s10686-017-9537-2}
  {\path{doi:10.1007/s10686-017-9537-2}}.

\bibitem{1992NIMPA.313..513G}
N.~{Gehrels}, {Instrumental background in gamma-ray spectrometers flown in low
  Earth orbit}, Nuclear Instruments and Methods in Physics Research A 313~(3)
  (1992) 513--528.
\newblock \href {https://doi.org/10.1016/0168-9002(92)90832-O}
  {\path{doi:10.1016/0168-9002(92)90832-O}}.

\bibitem{1999ApJ...520..124G}
D.~E. {Gruber}, J.~L. {Matteson}, L.~E. {Peterson}, G.~V. {Jung}, {The Spectrum
  of Diffuse Cosmic Hard X-Rays Measured with HEAO 1}, \apj 520~(1) (1999)
  124--129.
\newblock \href {http://arxiv.org/abs/astro-ph/9903492}
  {\path{arXiv:astro-ph/9903492}}, \href {https://doi.org/10.1086/307450}
  {\path{doi:10.1086/307450}}.

\bibitem{2002ApJ...576..188M}
D.~{McCammon}, R.~{Almy}, E.~{Apodaca}, et~al., {A High Spectral Resolution
  Observation of the Soft X-Ray Diffuse Background with Thermal Detectors},
  \apj 576~(1) (2002) 188--203.
\newblock \href {http://arxiv.org/abs/astro-ph/0205012}
  {\path{arXiv:astro-ph/0205012}}, \href {https://doi.org/10.1086/341727}
  {\path{doi:10.1086/341727}}.

\bibitem{wfibkg}
https://www.mpe.mpg.de/ATHENA-WFI/public/resources/background/WFI-MPE-ANA-Background-20150327.pdf.

\bibitem{1968ApJ...154.1011G}
L.~J. {Gleeson}, W.~I. {Axford}, {Solar Modulation of Galactic Cosmic Rays},
  \apj 154 (1968) 1011.
\newblock \href {https://doi.org/10.1086/149822} {\path{doi:10.1086/149822}}.

\bibitem{2018AnGeo..36..555F}
E.~{Frigo}, F.~{Antonelli}, D.~S.~S. {da Silva}, et~al., {Effects of solar
  activity and galactic cosmic ray cycles on the modulation of the annual
  average temperature at two sites in southern Brazil}, Annales Geophysicae
  36~(2) (2018) 555--564.
\newblock \href {https://doi.org/10.5194/angeo-36-555-2018}
  {\path{doi:10.5194/angeo-36-555-2018}}.

\bibitem{2004ApJ...614.1113M}
T.~{Mizuno}, T.~{Kamae}, G.~{Godfrey}, et~al., {Cosmic-Ray Background Flux
  Model Based on a Gamma-Ray Large Area Space Telescope Balloon Flight
  Engineering Model}, \apj 614~(2) (2004) 1113--1123.
\newblock \href {http://arxiv.org/abs/astro-ph/0406684}
  {\path{arXiv:astro-ph/0406684}}, \href {https://doi.org/10.1086/423801}
  {\path{doi:10.1086/423801}}.

\bibitem{Alcaraz:2000ks}
J.~Alcaraz, et~al., {Protons in near earth orbit}, Phys. Lett. B 472 (2000)
  215--226.
\newblock \href {http://arxiv.org/abs/hep-ex/0002049}
  {\path{arXiv:hep-ex/0002049}}, \href
  {https://doi.org/10.1016/S0370-2693(99)01427-6}
  {\path{doi:10.1016/S0370-2693(99)01427-6}}.

\bibitem{Alcaraz:2000vp}
J.~Alcaraz, et~al., {Cosmic protons}, Phys. Lett. B 490 (2000) 27--35.
\newblock \href {https://doi.org/10.1016/S0370-2693(00)00970-9}
  {\path{doi:10.1016/S0370-2693(00)00970-9}}.

\bibitem{Aguilar:2015ooa}
M.~Aguilar, et~al., {Precision Measurement of the Proton Flux in Primary Cosmic
  Rays from Rigidity 1 GV to 1.8 TV with the Alpha Magnetic Spectrometer on the
  International Space Station}, Phys. Rev. Lett. 114 (2015) 171103.
\newblock \href {https://doi.org/10.1103/PhysRevLett.114.171103}
  {\path{doi:10.1103/PhysRevLett.114.171103}}.

\bibitem{1972ZGeo...38..701M}
J.~{Moritz}, {Energetic protons at low equatorial altitudes.}, Zeitschrift fur
  Geophysik 38 (1972) 701--717.

\bibitem{2008AdSpR..41.1269P}
A.~N. {Petrov}, O.~R. {Grigoryan}, M.~I. {Panasyuk}, {Energy spectrum of proton
  flux near geomagnetic equator at low altitudes}, Advances in Space Research
  41~(8) (2008) 1269--1273.
\newblock \href {https://doi.org/10.1016/j.asr.2007.08.007}
  {\path{doi:10.1016/j.asr.2007.08.007}}.

\bibitem{2009AdSpR..43..654P}
A.~N. {Petrov}, O.~R. {Grigoryan}, N.~V. {Kuznetsov}, {Creation of model of
  quasi-trapped proton fluxes below Earth{\textquoteright}s radiation belt},
  Advances in Space Research 43~(4) (2009) 654--658.
\newblock \href {https://doi.org/10.1016/j.asr.2008.11.019}
  {\path{doi:10.1016/j.asr.2008.11.019}}.

\bibitem{2000SPIE.4140...32K}
E.~{Kendziorra}, T.~{Clauss}, N.~{Meidinger}, et~al., {Effect of low-energy
  protons on the performance of the EPIC pn-CCD detector on XMM-Newton}, in:
  K.~A. {Flanagan}, O.~H. {Siegmund} (Eds.), X-Ray and Gamma-Ray
  Instrumentation for Astronomy XI, Vol. 4140 of Society of Photo-Optical
  Instrumentation Engineers (SPIE) Conference Series, 2000, p. 32--41.
\newblock \href {https://doi.org/10.1117/12.409132}
  {\path{doi:10.1117/12.409132}}.

\bibitem{2003ITNS...50.2018L}
D.~H. {Lo}, J.~R. {Srour}, {Modeling of proton-induced ccd degradation in the
  chandra x-ray observatory}, IEEE Transactions on Nuclear Science 50~(6)
  (2003) 2018--2023.
\newblock \href {https://doi.org/10.1109/TNS.2003.820735}
  {\path{doi:10.1109/TNS.2003.820735}}.

\bibitem{2021ExA...tmp...12Q}
L.~{Qi}, G.~{Li}, Y.~{Xu}, et~al., {A preliminary design of the magnetic
  diverter on-board the eXTP observatory}, Experimental Astronomy (Mar. 2021).
\newblock \href {https://doi.org/10.1007/s10686-021-09707-x}
  {\path{doi:10.1007/s10686-021-09707-x}}.

\bibitem{2014PhRvL.113l1102A}
M.~{Aguilar}, D.~{Aisa}, A.~{Alvino}, et~al., {Electron and Positron Fluxes in
  Primary Cosmic Rays Measured with the Alpha Magnetic Spectrometer on the
  International Space Station}, \prl 113~(12) (2014) 121102.
\newblock \href {https://doi.org/10.1103/PhysRevLett.113.121102}
  {\path{doi:10.1103/PhysRevLett.113.121102}}.

\bibitem{2019PhRvL.122j1101A}
M.~{Aguilar}, L.~{Ali Cavasonza}, B.~{Alpat}, et~al., {Towards Understanding
  the Origin of Cosmic-Ray Electrons}, \prl 122~(10) (2019) 101101.
\newblock \href {https://doi.org/10.1103/PhysRevLett.122.101101}
  {\path{doi:10.1103/PhysRevLett.122.101101}}.

\bibitem{1994ApJ...436..769G}
R.~L. {Golden}, C.~{Grimani}, B.~L. {Kimbell}, et~al., {Observations of
  Cosmic-Ray Electrons and Positrons Using an Imaging Calorimeter}, \apj 436
  (1994) 769.
\newblock \href {https://doi.org/10.1086/174951} {\path{doi:10.1086/174951}}.

\bibitem{2008AdSpR..42.1523G}
O.~R. {Grigoryan}, M.~I. {Panasyuk}, V.~L. {Petrov}, et~al., {Spectral
  characteristics of electron fluxes at L \textless 2 under the Radiation
  Belts}, Advances in Space Research 42~(9) (2008) 1523--1526.
\newblock \href {https://doi.org/10.1016/j.asr.2007.12.009}
  {\path{doi:10.1016/j.asr.2007.12.009}}.

\bibitem{2013ITNS...60.3150G}
S.~{Granato}, R.~{Andritschke}, J.~{Elbs}, et~al., {Characterization of eROSITA
  PNCCDs}, IEEE Transactions on Nuclear Science 60~(4) (2013) 3150--3157.
\newblock \href {https://doi.org/10.1109/TNS.2013.2269907}
  {\path{doi:10.1109/TNS.2013.2269907}}.

\bibitem{2021A&A...647A...1P}
P.~{Predehl}, R.~{Andritschke}, V.~{Arefiev}, et~al., {The eROSITA X-ray
  telescope on SRG}, \aap 647 (2021) A1.
\newblock \href {http://arxiv.org/abs/2010.03477} {\path{arXiv:2010.03477}},
  \href {https://doi.org/10.1051/0004-6361/202039313}
  {\path{doi:10.1051/0004-6361/202039313}}.

\bibitem{1983ApJ...272..317L}
T.~P. {Li}, Y.~Q. {Ma}, {Analysis methods for results in gamma-ray astronomy.},
  \apj 272 (1983) 317--324.
\newblock \href {https://doi.org/10.1086/161295} {\path{doi:10.1086/161295}}.

\bibitem{2018ApJS..236...17V}
G.~{Vianello}, {The Significance of an Excess in a Counting Experiment:
  Assessing the Impact of Systematic Uncertainties and the Case with a Gaussian
  Background}, \apjs 236~(1) (2018) 17.
\newblock \href {http://arxiv.org/abs/1712.00118} {\path{arXiv:1712.00118}},
  \href {https://doi.org/10.3847/1538-4365/aab780}
  {\path{doi:10.3847/1538-4365/aab780}}.

\bibitem{xmmuhb}
”XMM-Newton Users Handbook”, Issue 2.17, 2019 (ESA: XMM-Newton SOC).

\bibitem{2020JHEAp..27...24L}
J.-Y. {Liao}, S.~{Zhang}, Y.~{Chen}, et~al., {Background model for the
  Low-Energy Telescope of Insight-HXMT}, Journal of High Energy Astrophysics 27
  (2020) 24--32.
\newblock \href {http://arxiv.org/abs/2004.01432} {\path{arXiv:2004.01432}},
  \href {https://doi.org/10.1016/j.jheap.2020.02.010}
  {\path{doi:10.1016/j.jheap.2020.02.010}}.

\end{thebibliography}





% \end{thebibliography}
\end{document}